\documentclass[pmlr,twocolumn,10pt]{jmlr} 

\usepackage{tablefootnote}
\usepackage{booktabs}
\usepackage{threeparttable}
\usepackage{subcaption}
\usepackage[colorinlistoftodos]{todonotes}
\usepackage{siunitx}
\usepackage[normalem]{ulem}

\usepackage[switch]{lineno}

\theorembodyfont{\upshape}
\theoremheaderfont{\scshape}
\theorempostheader{:}
\theoremsep{\newline}

\title[Removing eGFR Race Adjustment
]{Algorithmic Changes Are Not Enough: Evaluating the \\Removal of Race Adjustment from the eGFR Equation}

\author{%
\Name{Marika M. Cusick} \Email{marikacusick@stanford.edu}\\
\addr Stanford University School of Medicine
\AND
\Name{Glenn M. Chertow}\Email{gchertow@stanford.edu}\\
\addr Stanford University School of Medicine
\AND
\Name{Douglas K. Owens}\Email{owens@stanford.edu}\\
\addr Stanford University School of Medicine
\AND
\Name{Michelle Y. Williams}\Email{micwilliams@stanfordhealthcare.org}\\
\addr Stanford Health Care
\AND
\Name{Sherri Rose} \Email{sherrirose@stanford.edu}\\
\addr Stanford University School of Medicine
}

\begin{document}

\maketitle

\begin{abstract}
Changing clinical algorithms to remove race adjustment has been proposed and implemented for multiple health conditions. Removing race adjustment from estimated glomerular filtration rate (eGFR) equations may reduce disparities in chronic kidney disease (CKD), but has not been studied in clinical practice after implementation. Here, we assessed whether implementing an eGFR equation (CKD-EPI 2021) without adjustment for Black or African American race modified quarterly rates of nephrology referrals and visits within a single healthcare system, Stanford Health Care (SHC). Our cohort study analyzed 547,194 adult patients aged 21 and older who had at least one recorded serum creatinine or serum cystatin C between January 1, 2019 and September 1, 2023.  During the study period, implementation of CKD-EPI 2021 did not modify rates of quarterly nephrology referrals in those documented as Black or African American or in the overall cohort. After adjusting for capacity at SHC nephrology clinics, estimated rates of nephrology referrals and visits with CKD-EPI 2021 were 34 (95\% CI 29, 39) and 188 (175, 201) per 10,000 patients documented as Black or African American. If race adjustment had not been removed, estimated rates were nearly identical: 38 (95\% CI: 28, 53) and 189 (165, 218) per 10,000 patients. Changes to the eGFR equation are likely insufficient to achieve health equity in CKD care decision-making as many other structural inequities remain.
\end{abstract}

\paragraph*{Data and Code Availability}
We used electronic health record (EHR) data at Stanford Health Care (SHC) from the STAnford medicine Research data Repository (STARR) \citep{callahan2023stanford}. Because our data include protected health information, the data are not available to be shared publicly. Our code is available at: \href{https://github.com/StanfordHPDS/egfr_equation_shc}{github.com/StanfordHPDS/egfr\_equation\_shc}.

\paragraph*{Institutional Review Board (IRB)}
The Stanford University Institutional Review Board approved this research.

\section{Introduction}
\label{sec:intro}

Chronic kidney disease (CKD) currently affects more than 1 in 7 U.S. adults, about 37 million persons \citep{cdc2022}. Clinical management of CKD focuses on preventing kidney failure which requires dialysis or kidney transplantation, and preventing associated complications, including exceptionally high rates of cardiovascular disease. For racial and ethnic minorities, the burden of kidney failure is higher, with little to no improvement observed for several decades despite recognition of disparities since the 1980s \citep{vart2020national,hsu2003racial,rostand1982racial}. Black or African American and Hispanic patients are at least 3-fold and 1.5-fold more likely to progress to kidney failure in comparison to non-Hispanic white patients \citep{desai2019ckd,united20202020}. Suboptimal management of mild-to-moderate CKD in the primary care setting, including delays in referrals and visits to nephrology, have contributed to disparities in CKD and kidney failure outcomes \citep{navaneethan2008systematic}.
Primary care providers typically rely on the estimated glomerular filtration rate (eGFR) equation to gauge severity of CKD and inform CKD care decisions. As eGFR values decrease, patients are classified into more severe CKD stages. While several such equations were developed, the two most widely adopted equations (the 4-variable MDRD Study equation and the CKD-EPI 2009 equation) incorporated age, sex, Black versus non-Black race, and serum creatinine. The MDRD and CKD-EPI 2009 equations include coefficients that increase eGFR estimates 21\% and 16\% higher, for patients administratively documented as Black or African American \citep{levey2009new,levey1999more,eneanya2019reconsidering}.

The inclusion of race in clinical algorithms, such as the eGFR equation, propagates racial bias in decision-making \citep{vyas2020hidden}. Calls for a non-race-adjusted eGFR equation led to an effort supported by two professional organizations and a novel eGFR equation (CKD-EPI 2021) that was estimated without race. In retrospective validation studies, the CKD-EPI 2021 eGFR equation underpredicted measured GFR, the gold standard, for Black patients, yet overpredicted measured GFR for non-Black patients \citep{inker2021new}. Many health care systems subsequently implemented and deployed CKD-EPI 2021 into clinical care \citep{miller2022national,genzen2022reported,genzen2023update}. The literature on algorithmic bias and fairness has proposed additional potential changes beyond removing race adjustment for health care algorithms, including fairness constraints in the loss function, but such changes have not been deployed in practice for CKD \citep{chen2021ethical}.

Implementation of a new eGFR equation without race adjustment has the potential to reduce downstream disparities in CKD and kidney failure, particularly for Black or African American patients, by lowering GFR estimates and promoting early detection and treatment of CKD \citep{ahmed2021examining,gregg2022effects,ku2022comparison,ghuman2022impact}. However, racial disparities in CKD cannot be explained by eGFR-guided CKD decision-making alone, as other factors, including higher prevalence of other comorbid conditions (diabetes, hypertension), lower socioeconomic status, poorer housing and neighborhood conditions, and other systemically racist policies and practices, contribute to CKD inequities \citep{norton2016social}.
      
To our knowledge, prospective assessments of the new equation’s effect on CKD care decision-making and health outcomes have not been reported. Herein, we estimate the effects of implementing CKD-EPI 2021 on nephrology referrals and visits for patients within a single health care system. 
\section{Methods}

\subsection{Data, Study Population, and Measures}

In this cohort study, we identified all adult patients aged 21 and older with at least one serum creatinine or serum cystatin C at an SHC hospital or clinic recorded in STARR during our study period: January 1, 2019 through September 1, 2023. At SHC, the CKD-EPI 2021 eGFR equation was implemented on December 1, 2021 in Epic EHR systems. eGFR was computed automatically for chemistry panels and point of care services, requiring no behavior change from SHC providers. We manually validated the implementation and uptake of CKD-EPI 2021 in STARR by computing the proportion of measurements that rely on the new equation (\appendixref{apd:methods1}, \tableref{tab:etab1}). For eGFR measurements recorded after December 1, 2021, we compared differences in eGFR values and CKD stages as calculated with CKD-EPI 2021 and CKD-EPI 2009, the eGFR formula most commonly used at SHC prior to the implementation of CKD-EPI 2021 (\appendixref{apd:methods2}, \tableref{tab:etab1}).

Our outcomes were quarterly rates of nephrology referrals and visits (\appendixref{apd:methods3}). Nephrology referrals are often prerequisites to nephrology visits. A given patient would have no more than one nephrology referral, whereas nephrology visits can occur repeatedly. We defined quarters to align with the date of CKD-EPI 2021 implementation: December – February, March – May, June – August, and September – November. Rates were normalized by number of patients with any visit at SHC per quarter during our study period (\figureref{fig:efig1,fig:efig2,fig:efig3}). 

\subsection{Statistical Analyses}

We estimated the effect of implementing the CKD-EPI 2021 equation on quarterly rates of nephrology referrals and visits using an interrupted time series (ITS) study design estimated by Poisson regression, a common choice for count outcomes \citep{bernal2017interrupted}. In our ITS formulation, the underlying time series trend of SHC nephrology referrals and visit rates was interrupted by the implementation of CKD-EPI 2021.

Our main results assumed the eGFR equation change had a gradual effect on rates of nephrology referrals and visits, which we formalized in an impact model with a temporary slope change followed by a level change in our rates of interest. We expected gradual changes after the eGFR equation change given use of the CKD-EPI 2021 equation steadily increased across SHC hospitals and clinics in the months following December 1, 2021. We assumed the period associated with the temporary slope change was nine months, which corresponds to when usage of the CKD-EPI 2021 equation was at least 90\% across SHC hospitals and clinics (\figureref{fig:efig4}). The unadjusted ITS regression was given by:
\begin{align}
Y_t & = \beta_0 + \beta_1T_t + \beta_2A_t, \nonumber\\
& Y_t: \text{Quarterly rate at time } t,\nonumber\\
&T_t: \text{Quarters elapsed at time } t,\nonumber\\
&A_t: \text{CKD-EPI 2021 at time } t,\nonumber\end{align}
where $t=\{1, ...,18\}$ was measured in quarters. The indicator for CKD-EPI 2021, $A_t$, was 0 prior to CKD-EPI 2021 implementation, between 0 and 1 during the 9-month temporary period, and 1 afterward. The temporary period values were directly tied to the temporary slope change impact model with $A_t$ values of 0.25, 0.50, 0.75 at 3, 6, and 9 months post-implementation. The effect of interest, $\beta_2$, represents the impact of implementing CKD-EPI 2021.
We conducted the ITS subgroup analyses for patients documented as Black or African American or not Black or African American as well as analyses for the overall SHC population (\appendixref{apd:methods4}).

We reported estimated quarterly rates of nephrology referrals and visits after December 1, 2021, with and without implementation of the CKD-EPI 2021 equation. Changes to rates from the implementation of the CKD-EPI 2021 equation were estimated by unadjusted and adjusted rate ratios (RRs with 95\% CIs). To account for capacity at SHC nephrology clinics, a possible time-varying covariate, we adjusted for the median number of days from nephrology referral to visit at a nephrology clinic (\appendixref{apd:methods5}, \tableref{tab:etab2,tab:etab3}, \figureref{fig:efig5,fig:efig6,fig:efig7}). The adjusted ITS regression was given by: $Y_t = \beta_0 + \beta_1T_t + \beta_2A_t + \beta_3X_t,$ where $X_t$ was median days from referral to visit at time $t$.

In sensitivity analyses, we tested varying temporary slope change periods (3, 6, and 12 months), an alternative impact model (assuming an immediate level and slope change on rates of nephrology referrals and visits), and another indicator for capacity: average number of active providers at SHC nephrology clinics (defined as providers with at least one visit at a SHC nephrology clinic). We considered the inclusion of other possible time-varying covariates, such as SHC demographics (average age, proportion of documented female patients) and number of patients in our cohort with common comorbidities (diabetes and hypertension). Finally, we tested our regressions assuming rates of nephrology referrals and visits followed seasonal patterns, as patient counts at SHC can follow seasonal patterns from infectious diseases (\appendixref{apd:methods6}) \citep{fisman2007seasonality}.

We stored data on a secure Google Cloud Storage server and processed using BigQuery SQL workplace. We conducted all analyses in Python version 3.9.1. 

\begin{table*}[h]
    \floatconts {tab:cohortstudy}%
      {\caption{Characteristics of adult patients with at least one serum creatinine or cystatin C recorded at Stanford Health Care hospitals and clinics}}%
    {%
    \begin{tabular}{llr}
    \bfseries Characteristic &  & \\
    \hline
     \abovestrut{2.2ex}
      Age$^*$ (mean (SD))&  & 48 (18) \\   
    \rule{0pt}{3ex}Documented sex ($n$ (\%))& Female & 298,680 (55\%) \\
     & Male & 248,345 (45\%) \\
     & Unknown & 169 ($<$1\%) \\
     \rule{0pt}{3ex}Documented race$^{**}$ ($n$ (\%))& American Indian or Alaska Native & 2,257 ($<$1\%) \\
     & Asian & 121,673 (22\%) \\
     & Black or African American & 24,373 (5\%) \\
     & Native Hawaiian or Other Pacific Islander & 5,555  (1\%) \\
     & White & 271,604 (50\%) \\
     & Additional group & 105,545 (20\%) \\
     & Decline to State & 13,686 (3\%) \\
     & Unknown & 24,727 (5\%) \\
     \rule{0pt}{3ex}Documented ethnicity$^{**}$ ($n$ (\%)) & Hispanic/Latino & 82,639 (15\%) \\
     & Not Hispanic/Latino &  435,138 (80\%)\\
     & Decline to State & 17,935 (3\%) \\
     & Unknown & 26,041 (5\%) \\
   \hline
    \end{tabular}
}
\begin{tablenotes}\footnotesize
\item[*] $^*$ At first observed visit at Stanford Health Care
\item[*] $^{**}$ Percentages do not sum to 1 as some patients were documented in multiple categories

\end{tablenotes}
\end{table*}

\section{Results}
\subsection{Study Population and Outcomes}

A total of 547,194 adult patients had at least one serum creatinine or cystatin C value at SHC hospitals and clinics during our study period. Mean age was 48 years and 298,680 (55\%) were female. Documented race and ethnicity characteristics were: 2,257 ($<$1\%) American Indian or Alaska Native, 121,673 (22\%) Asian, 24,373 (5\%) Black or African American, 5,555 (1\%) Native Hawaiian or Other Pacific Islander, 271,604 (50\%) white, and 105,545 (20\%) additional group as well as 82,639 (15\%) Hispanic/Latino  (\tableref{tab:cohortstudy}).

\begin{table*}[htbp]
    \floatconts {tab:observedrates}%
      {\caption{Mean observed quarterly rates for nephrology referrals and visits}}%
    {%
    \begin{tabular}{ llll}
    \bfseries  & \bfseries   & \bfseries Black or  &  \bfseries Not Black or  \\
    \bfseries  & \bfseries Overall & \bfseries African American$^*$ &  \bfseries African American$^*$ \\
    \hline\abovestrut{2.2ex}
     Rate of nephrology referrals  & 20 & 33 & 19\\
     (per 10,000 patients) & & & \\
    \rule{0pt}{3ex}Rate of nephrology visits  & 99 & 167 & 96\\
    (per 10,000 patients) & & & \\
    \rule{0pt}{3ex}Patients with any visit at SHC & 263,742 & 11,711 & 252,067\\
     \belowstrut{0.2ex} 
    \rule{0pt}{3ex}Median time from referral to visit & 57 & 58 & 57\\\hline
    \end{tabular}
}
\begin{tablenotes}\footnotesize
\item[*] $^*$ Documented race
\end{tablenotes}
\end{table*}

In our cohort, 9,329 (2\%) and 10,676 (2\%) patients had nephrology referrals and visits, respectively, where, for example, visits include patients who had external referrals that are not documented in the EHR. Of those with referrals and visits, 687 (7\%) and 738 (7\%) were observed in patients documented as Black or African American. During the study period, the average observed quarterly rates of nephrology referrals and visits were 20 and 99 per 10,000 patients. Among patients documented as Black or African American and not Black or African American, average observed quarterly rates for nephrology referral and visits were 33 and 167 per 10,000 patients and 19 and 96 per 10,000 patients, respectively (\tableref{tab:observedrates}). The median time from nephrology referral to visit at a nephrology clinic increased from 32 days in 2019 to 100 days in 2023 (\tableref{tab:etab2}, \figureref{fig:efig5,fig:efig6,fig:efig7}).

\subsection{Main Analyses}

When compared to CKD-EPI 2009, CKD-EPI 2021 produced the largest changes in eGFR values for patients documented as Black or African American, for whom eGFR values decreased on average by 10\%, and 18\% of measurements were assigned to more severe CKD stages (\tableref{tab:etab4,tab:etab5,tab:etab6}). For those not documented as Black or African American, eGFR values increased by 5\% on average and 12\% were assigned to less severe CKD stages (\tableref{tab:etab4,tab:etab7,tab:etab8}). The majority (58\%) of changes in CKD stage assignment were between CKD stages G1 and G2, the least severe stages of CKD (\tableref{tab:etab9,tab:etab10}). 

\begin{figure}
\floatconts {fig:fig1}
  {\caption{Observed and estimated quarterly rates (with shaded 95\% CIs) of nephrology referrals with CKD-EPI 2021 and race-adjusted eGFR from interrupted time series regression during study period }}
    {\subfigure[Black or African American$^*$]{\label{fig:figure1a}
    \includegraphics[width=0.45\textwidth]{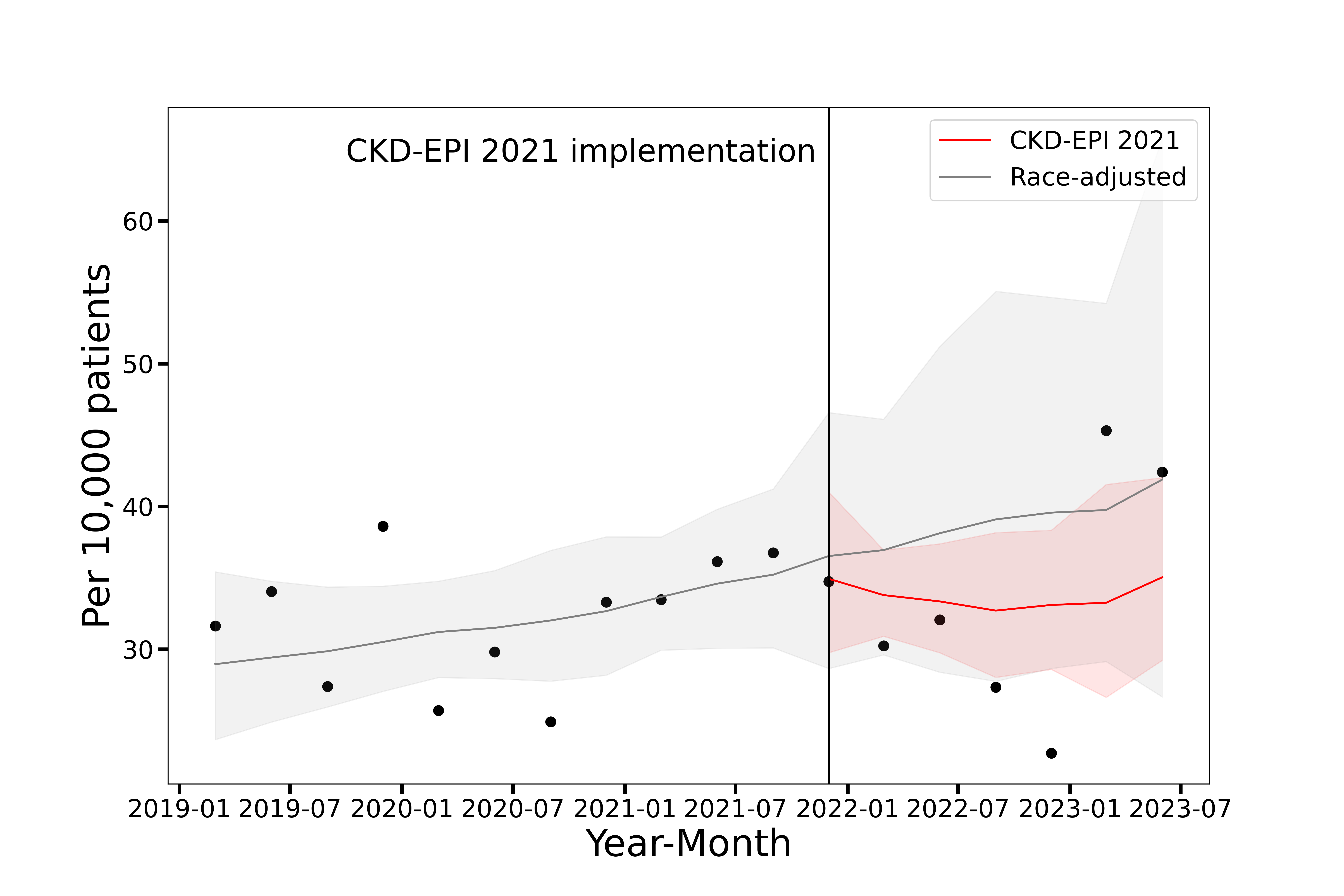}}
    \vspace{1em} 

    \subfigure[Not Black or African American$^*$]{\label{fig:figure1b}
    \includegraphics[width=0.45\textwidth]{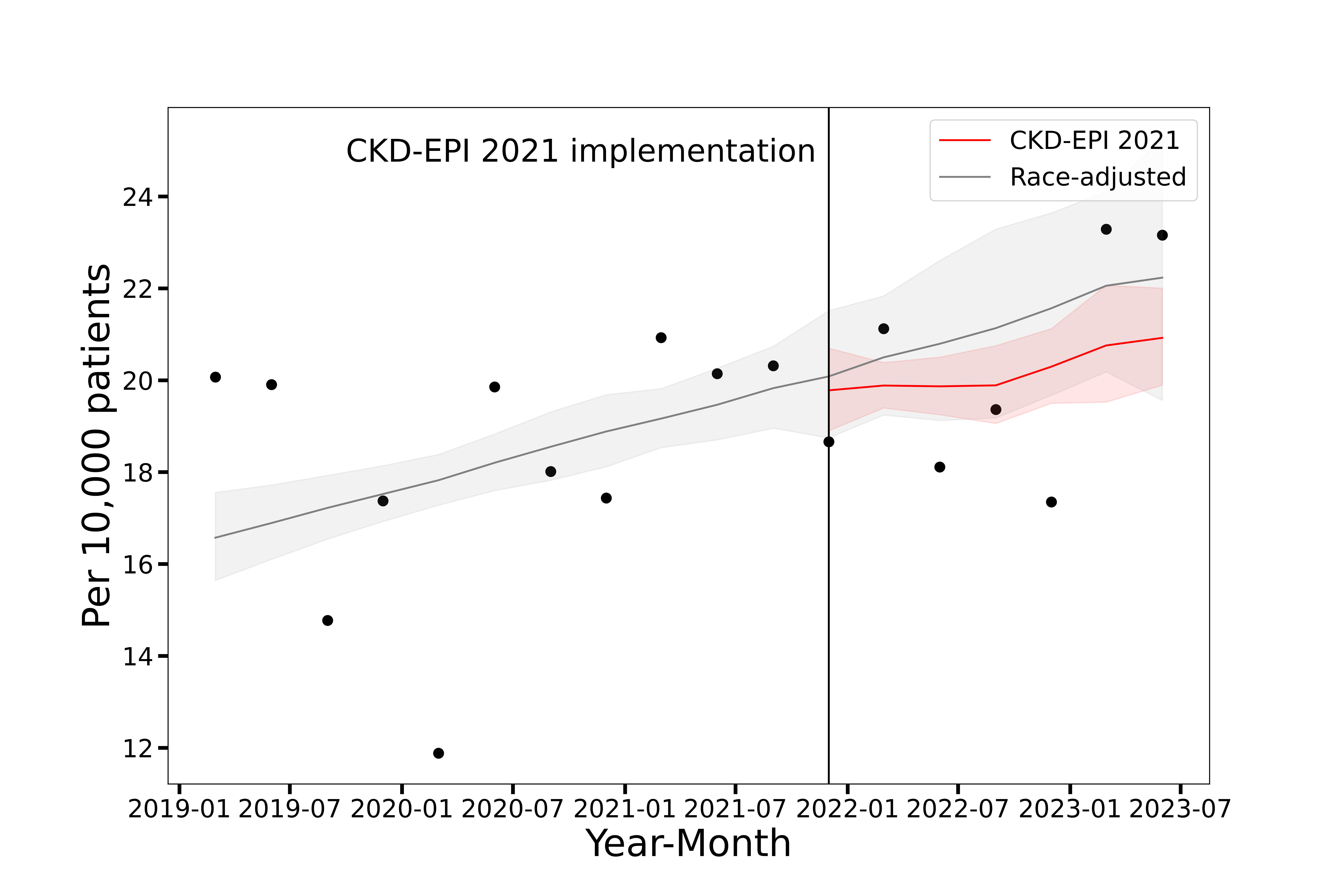}}
    \vspace{1em} 

    \subfigure[Overall]{\label{fig:figure1c}
    \includegraphics[width=0.45\textwidth]{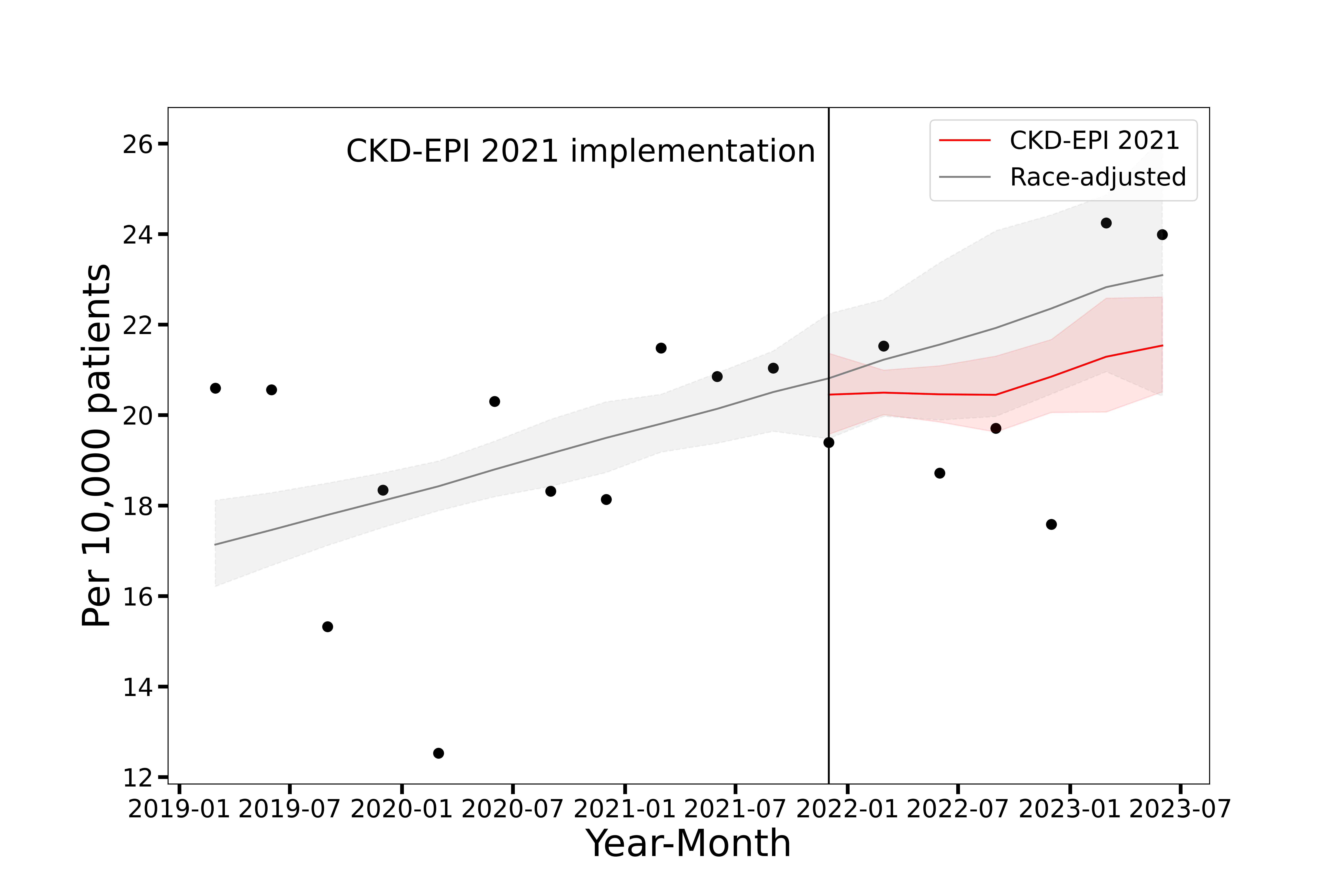}}
    \vspace{1em} 
    }
    \begin{tablenotes}\footnotesize
\item[*] $^*$ Documented race
\end{tablenotes}
\end{figure}

\begin{figure}
\floatconts{fig:fig2}
  {\caption{Observed and estimated quarterly rates (with shaded 95\% CIs) of nephrology visits with CKD-EPI 2021 and race-adjusted eGFR from interrupted time series regression during study period }}
    {\subfigure[Black or African American$^*$]{\label{fig:figure2a}
    \includegraphics[width=0.45\textwidth]{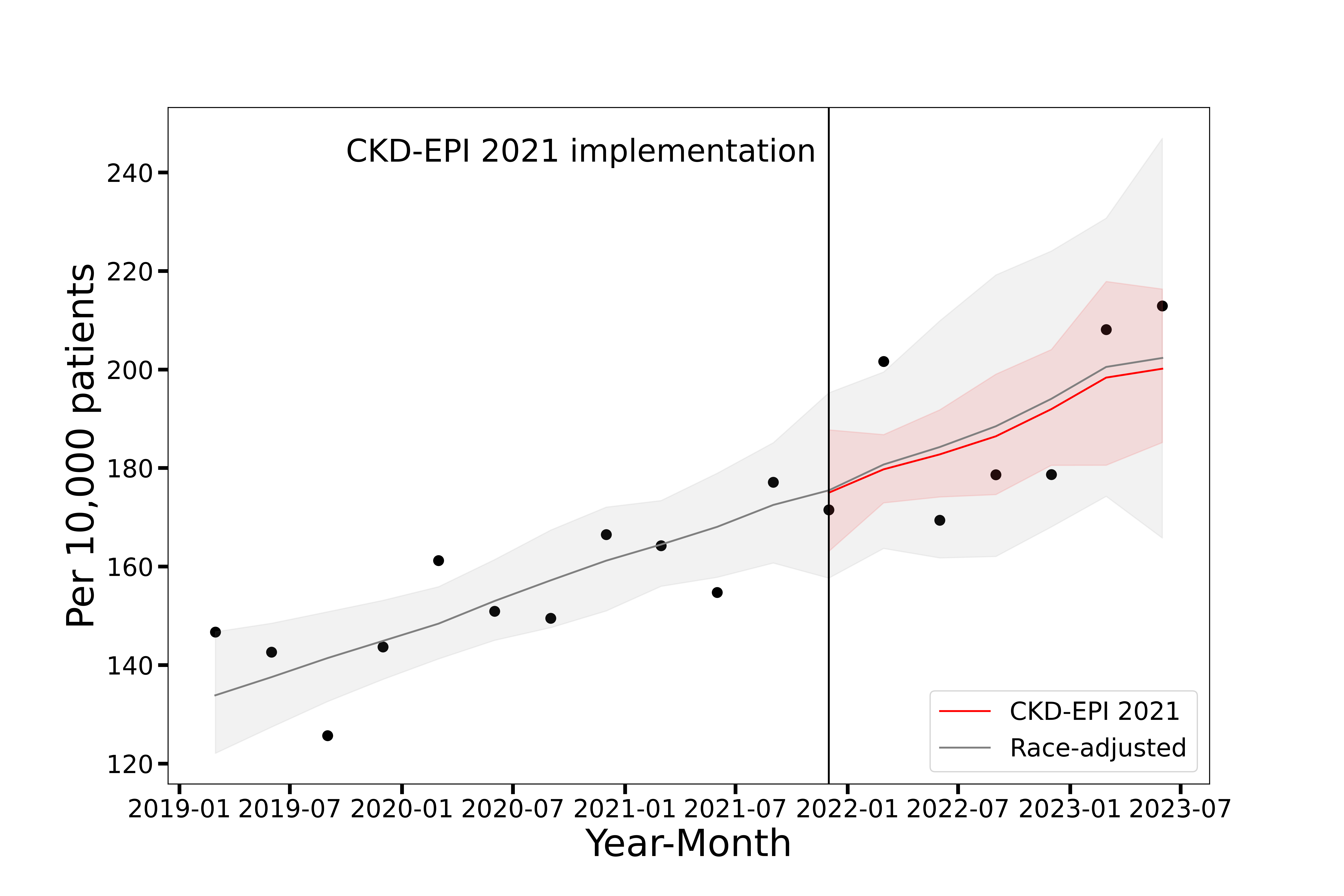}}
    \vspace{1em} 

    \subfigure[Not Black or African American$^*$]{\label{fig:figure2b}
    \includegraphics[width=0.45\textwidth]{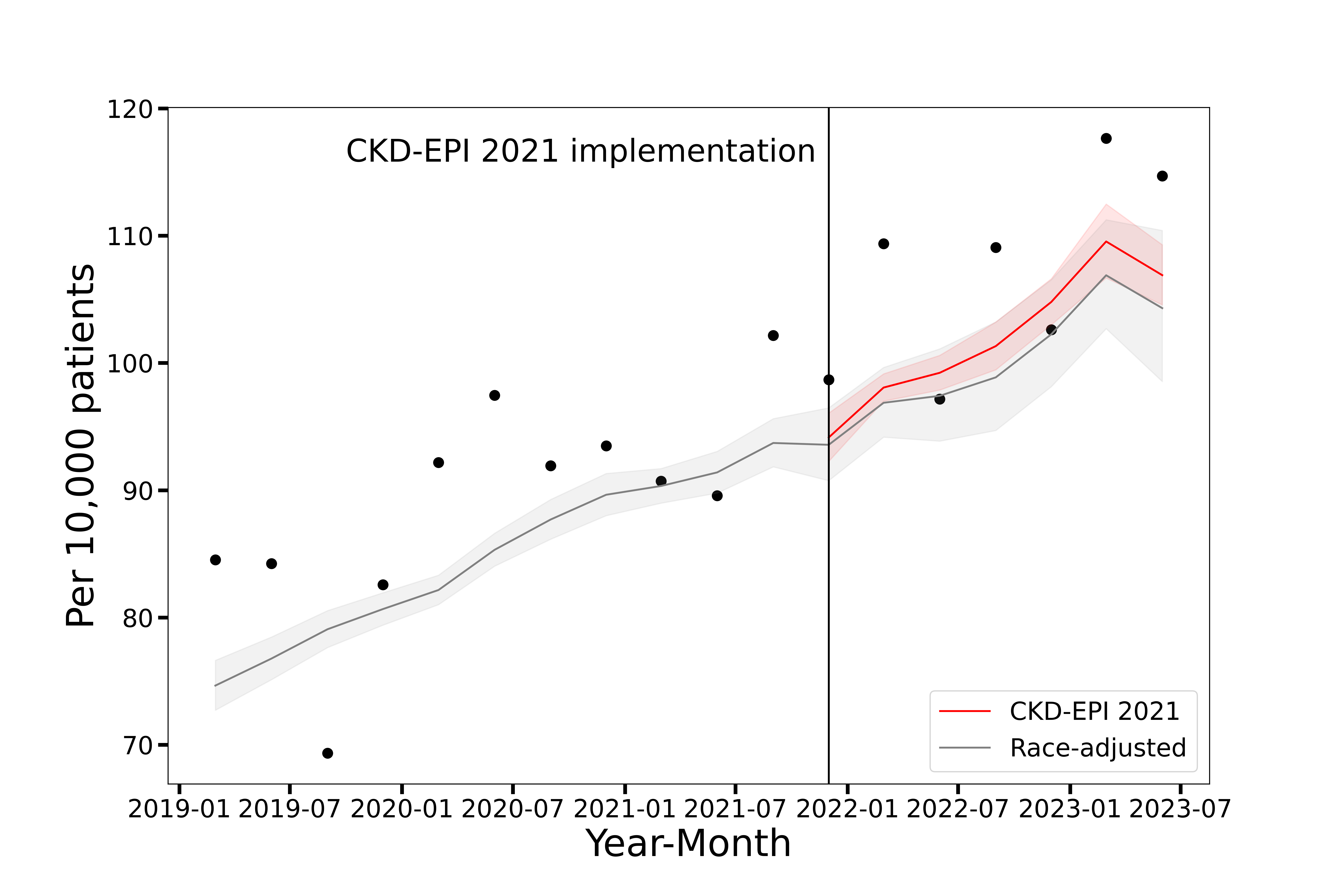}}
    \vspace{1em} 

    \subfigure[Overall]{\label{fig:figure2c}
    \includegraphics[width=0.45\textwidth]{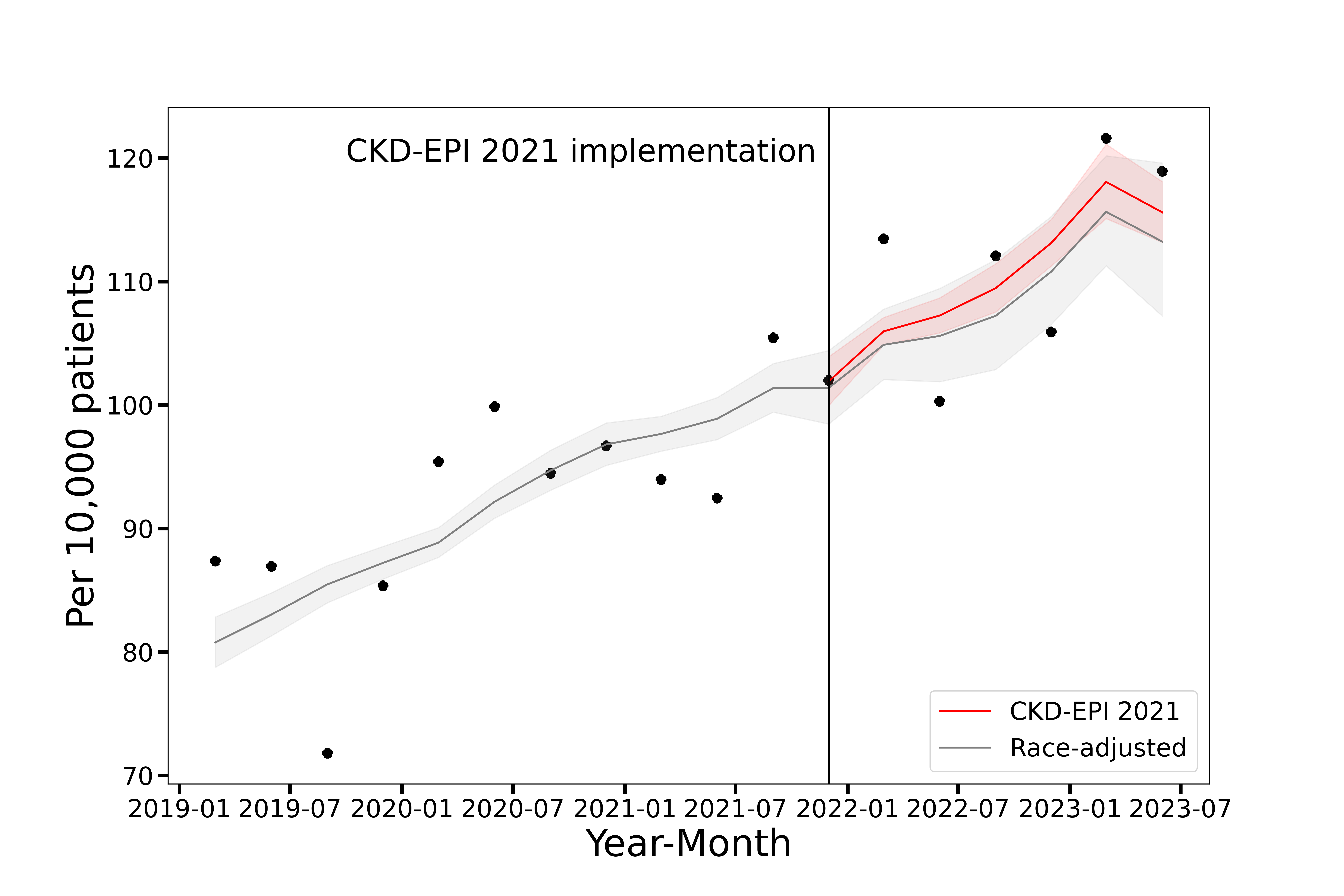}}
    \vspace{1em} 
    }
    \begin{tablenotes}\footnotesize
\item[*] $^*$ Documented race
\end{tablenotes}
\end{figure}

Estimated quarterly rates of nephrology referrals and visits did not differ following implementation of the CKD-EPI 2021 eGFR equation among studied subgroups (\figureref{fig:fig1,fig:fig2}). After implementation of the CKD-EPI 2021 equation, the estimated quarterly rate of nephrology referrals was 34 (95\% CI: 29, 39 per 10,000 patients documented as Black or African American. If the CKD-EPI 2021 eGFR equation had not been implemented (i.e., race-adjusted), the estimated quarterly rate was 38 (95\% CI: 28, 53) per 10,000 patients documented as Black or African American. With and without the implementation of CKD-EPI 2021, quarterly rates of nephrology visits did not differ: 188 (95\% CI: 175, 201) and 189 (95\% CI: 165, 218) per 10,000 patients documented as Black or African American. For patients documented as not Black or African American, estimated rates of nephrology referrals and visits were 20 (95\% CI: 19, 21) and 102 (95\% CI: 100, 104) per 10,000 patients with CKD-EPI 2021 and 21 (95\% CI: 19, 23) and 100 (95\% CI: 96, 104) per 10,000 patients without CKD-EPI 2021  (\tableref{tab:itsresults}).

\begin{table*}[htbp]
    \floatconts {tab:itsresults}%
      {\caption{Rate ratios (RR) of implementing CKD-EPI 2021 and estimated quarterly rates of nephrology referrals and visits under CKD-EPI 2021 and race-adjusted eGFR with confidence intervals (95\% CIs) from interrupted time series regression for overall cohort and subgroups (Black or African American, not Black or African American)}}%
    {%
    \begin{tabular}{ llll}
    \bfseries  & \bfseries   & \bfseries Black or  &  \bfseries Not Black or  \\
    \bfseries  & \bfseries Overall & \bfseries African American$^*$ &  \bfseries African American$^*$ \\
    \hline\abovestrut{2.2ex}
    \textbf{Nephrology referral rates} & & & \\ 
    \rule{0pt}{3ex}Unadjusted RR & 0.93 (0.85, 1.02) & 0.85 (0.61, 1.20) & 0.94 (0.85, 1.03)\\
    (CKD-EPI 2021)  & & & \\
   \rule{0pt}{3ex}Adjusted RR  & 0.93 (0.84, 1.04) & 0.84 (0.57, 1.23) & 0.94 (0.84, 1.05)\\
    (CKD-EPI 2021)  & & & \\
    \rule{0pt}{3ex}Estimated rate per 10,000 people & 21 (20, 22) & 34 (29, 39) & 20 (19, 21)\\
    (CKD-EPI 2021)  & & & \\
    \rule{0pt}{3ex}Estimated rate per 10,000 people & 22 (20, 24) & 38 (28, 53) & 21 (19, 23)\\
    (race adjusted)  & & & \\
    \rule{0pt}{4ex}\textbf{Nephrology visit rates} & & & \\  
    \rule{0pt}{3ex}Unadjusted RR  & 0.99 (0.95, 1.03) & 0.98 (0.84, 1.14) & 0.99 (0.95, 1.04)\\
    (CKD-EPI 2021)  & & & \\
    \rule{0pt}{3ex}Adjusted RR  & 1.02 (0.97, 1.07) & 0.99 (0.83, 1.17) & 1.03 (0.98, 1.08)\\
    (CKD-EPI 2021)  & & & \\
   \rule{0pt}{3ex}Estimated rate per 10,000 people & 110 (108, 112) & 188 (175, 201) & 102 (100, 104)\\
    (CKD-EPI 2021)  & & & \\
    \belowstrut{0.2ex}
    \rule{0pt}{3ex}Estimated rate per 10,000 people & 108 (104, 113) & 189 (165, 218) & 100 (96, 104)\\
    (race adjusted)  & & & \\ \hline
    \end{tabular}
}
\begin{tablenotes}\footnotesize
\item[*] $^*$ Documented race
\end{tablenotes}
\end{table*}

For patients documented as Black or African American, the unadjusted RRs of the CKD-EPI 2021 equation implementation on nephrology referrals and visits was 0.85 (95\% CI: 0.61, 1.20, p-value: 0.37) and 0.98 (95\% CI:0.84, 1.14, p-value: 0.78). After adjusting for capacity at SHC nephrology clinics, the corresponding RRs were 0.84 (95\% CI: 0.57, 1.23, p-value: 0.37) and 0.99 (95\% CI: 0.83, 1.17, p-value: 0.90). For those documented as not Black or African American, adjusted RRs of the CKD-EPI 2021 equation implementation on nephrology referrals and visits were 0.94 (95\% CI: 0.84, 1.05, p-value: 0.28) and 1.03 (95\% CI: 0.98, 1.08, p-value: 0.32) (\tableref{tab:itsresults}). 

When examining the overall cohort, estimated quarterly rates of nephrology referrals and visits did not change after implementing the CKD-EPI 2021 equation. If CKD-EPI 2021 were not implemented, estimated rates were 22 (95\% CI: 20, 24) and 108 (95\% CI: 104, 113) per 10,000 patients for nephrology referrals and visits, respectively. With implementation of CKD-EPI 2021, estimated rates were 21 (95\% CI: 20, 22) and 110 (95\% CI: 108, 112) per 10,000 patients, respectively (\tableref{tab:itsresults}). 

\subsection{Sensitivity Analyses}

Our results were robust across sensitivity analyses. There were no statistically significant differences in the quarterly rates of nephrology referrals and visits under alternative temporary slope change periods, impact models, adjustment with another potential time-varying covariate (active providers at SHC nephrology clinics), and seasonal adjustments (\tableref{tab:etab11,tab:etab12,tab:etab13,tab:etab14,tab:etab15,tab:etab16}). The additional potential time-varying covariates also did not appear to be strongly time-varying over our study period (\figureref{fig:efig8,fig:efig9,fig:efig10,fig:efig11}).

\section{Discussion}

In the two years following implementation of the CKD-EPI 2021 equation–-a new eGFR equation without race adjustment–-there were no changes in referral or visit rates among patients with mild-to-moderate CKD within a single healthcare system. When comparing the CKD-EPI 2021 equation without race adjustment to previously utilized eGFR equations, eGFR estimates were consistently lower for patients documented as Black or African American, resulting in a higher proportion of patients documented as Black or African American being classified in more severe CKD stages.  Despite these differences, we observed no change to nephrology referrals and visits after implementation of the new equation.  

There are a number of reasons that may be underlying the results seen here where we did not observe changes in rates of nephrology referrals and visits. First, CKD care decision-making (including referral to nephrology) relies on a myriad of factors, including eGFR, albuminuria, and presence of comorbid conditions including diabetes, obesity, hypertension, and cardiovascular disease. Second, changes to eGFR after the implementation of CKD-EPI 2021 may not be large enough to meaningfully influence system-wide nephrology referrals and visits. The majority of changes were in the earliest stages of CKD, where the eGFR equations show the poorest concordance with measured GFR \citep{inker2021new}. Third, CKD-EPI 2021 produced the largest changes in eGFR values for patients documented as Black or African American, who only made up 5\% of our study population. Our results may be limited by small sample sizes, and health systems with a larger proportion of patients documented as Black or African American may observe different results. 

More importantly, racial disparities in CKD affecting Black or African American patients cannot be attributed to eGFR-guided nephrology referral patterns alone. Previous studies have found that Black or African American patients had higher rates of nephrology care, anti-hypertensive medication usage, and albuminuria testing compared to non-Hispanic White patients \citep{chu2021trends,suarez2018racial}. This suggests that disparities in CKD are also driven by factors outside the healthcare delivery process, such as social determinants of health and structural racism \citep{norton2016social,crews2022designing,boulware2021seen}. Prior research cites lower socioeconomic status \citep{norris2008race,norris2021social}, lower rates of health insurance coverage and access to usual medical care \citep{jurkovitz2013association,evans2011race}, increased stress from racial discrimination \citep{bruce2015stress,bruce2009social,camelo2018racial}, and poorer environmental and neighborhood conditions \citep{volkova2008neighborhood,gutierrez2015contextual}, as contributors to inequities in CKD. 

While the inclusion of race adjustment in clinical equations contributes to racial bias, changes to the eGFR equation are insufficient to tackle social factors and structural inequities. A focus on equations should not divert efforts from research and interventions that aim to tackle these structural causes of health and health care disparities in kidney disease. 

Use of the CKD-EPI 2021 eGFR equation has been hypothesized to influence clinical decisions beyond nephrology referrals and visits, such as prescription medication eligibility, acute kidney injury treatment, vascular access referral, and kidney transplant eligibility \citep{ahmed2021examining,ghuman2022impact,uzendu2023implications}. We did not examine these outcomes due to sample size and definition limitations. Our analyses focused on nephrology referrals and visits, as we did not expect meaningful changes in CKD and kidney failure outcomes within a two-year period. Future research could evaluate these outcomes with a longer follow-up period after the implementation of CKD-EPI 2021.

There are limitations to our study. First, our study relies on EHR data in a single health system, which are limited to clinical encounters within SHC. Second, the reliability of race and ethnicity information in EHR data are unclear. Previous studies have reported lack of concordance between self-reported and administratively recorded race and ethnicity information in other health systems \citep{polubriaginof2019challenges,hamilton2009concordance}. Third, we categorized all other racial and ethnic groups as not documented as Black or African American in our subgroup analyses, which does not allow us to explore whether nephrology referrals differed among additional minoritized groups over the study period. Fourth, the COVID-19 pandemic occurred during our study period, which broadly influenced clinical care, including referral patterns and provider workload. We aimed to address this by adjusting for changes to clinical demand over the course of the study period. However, it is possible that other factors related to the COVID-19 pandemic influenced our results. Finally, there may be other unknown policy changes related to CKD care at SHC that impact rates of nephrology referrals and visits. 

\section{Conclusion}

Removing race adjustment from the eGFR equation has the potential to reduce disparities in nephrology care for Black or African American patients. After two years of follow-up in a single health system, implementation of the CKD-EPI 2021 eGFR equation did not result in changes to nephrology referrals or visits for patients documented as Black or African American. Mitigating racial and ethnic disparities in CKD care will require a continued focus on social and structural causes of inequities across all aspects of health care in the US. Algorithmic changes will not be enough.

\acks{This work was funded by a grant from the Stanford Impact Labs. \newline

\noindent We include the following acknowledgement text as guided by the Office of the Senior Associate Dean for Research at Stanford School of Medicine (\href{https://starr.stanford.edu/resources/faq}{starr.stanford.edu/resources/faq}): 
\begin{quote}
``This research used data or services provided by STARR, “STAnford medicine Research data Repository,” a clinical data warehouse containing live Epic data from Stanford Health Care, the Stanford Children’s Hospital, the University Healthcare Alliance and Packard Children's Health Alliance clinics and other auxiliary data from Hospital applications such as radiology PACS. STARR platform is developed and operated by Stanford Medicine Research Technology team and is made possible by Stanford School of Medicine Research Office.'' 
\end{quote}
We also would like to acknowledge Jeremy Goldhaber-Fiebert, Nakaya Frazier, Kelvin Nguyen, and Joshua Salomon for providing feedback and insights on this research. }

\bibliography{References}

\appendix
\renewcommand{\thefigure}{B\arabic{figure}}
\setcounter{figure}{0} 
\section{Appendix Methods}\label{apd:first}
\renewcommand{\thetable}{B\arabic{table}}
\setcounter{table}{0} 

\subsection{Manual Validation of SHC eGFR Equation Change}\label{apd:methods1}
On November 12, 2021, SHC announced that they would be implementing the new eGFR equation (CKD-EPI 2021) that no longer adjusts for race. Because our analysis relied on the implementation of CKD-EPI 2021, we conducted a manual validation using STARR data to confirm changes in usage of eGFR equation types over time across the SHC health system. 

We identified all unique serum creatinine measurements (4,103,122) and eGFR measurements (6,030,862) at SHC since the beginning of our study period. Because serum creatinine and eGFR measurements are often ordered simultaneously, we matched serum creatinine and eGFR measurements according to unique visit identifiers and measurement date and time. We identified at least one eGFR measurement match for 84\% of the serum creatinine measurements. 

Using serum creatinine values, age at measurement, sex, and race (administratively documented as Black or African American), we manually computed eGFR values according to the following commonly used equations: CKD-EPI 2021, CKD-EPI 2009 (adjusted for Black or African American), CKD-EPI 2009 (adjusted for those not Black or African American), MDRD (adjusted for Black or African American), and MDRD (adjusted for those not Black or African American). We computed the absolute differences between the recorded eGFR values and manually computed eGFR values, and we assumed the eGFR equation type to be the one with the smallest absolute difference in eGFR values. Of the assigned eGFR equation types, 8\% and 3\% had an absolute difference in eGFR values greater than 1 mL/min/1.73m2 and 2 mL/min/1.73m2. 

Prior to the eGFR equation change at SHC on December 1, 2021, most visits (86\%) with an eGFR measurement were reliant on the CKD-EPI 2009 equation, which had race adjustment according to whether patients are documented as Black or African American. However, after December 1, 2021, most visits (87\%) relied on the CKD-EPI 2021 equation, which no longer adjusts for race (\tableref{tab:etab1}). 

\subsection{Differences in eGFR Values and CKD Stages When Calculated with CKD-EPI 2021 and CKD-EPI 2009}\label{apd:methods2}

For all eGFR values recorded after December 1, 2021, we compared eGFR values calculated with CKD-EPI 2009 and CKD-EPI 2021 eGFR equations. eGFR values calculated with CKD-EPI 2009 represent the counterfactual scenario in which the eGFR equation did not change and continued to have race adjustment. 

We identified a total of 1,446,144 eGFR measurements at SHC, of which 69,747 (5\%) corresponded to patients documented as Black or African American. Average percent change between eGFR values calculated with CKD-EPI 2009 and CKD-EPI 2021 was -10\% for those documented as Black or African American and 5\% for those not documented as Black or African American (\tableref{tab:etab4}). 

Among the 69,747 eGFR measurements corresponding to patients documented as Black or African American, a total of 12,456 (18\%) measurements had a different CKD stage depending on whether eGFR was computed with CKD-EPI 2009 or CKD-EPI 2021. The proportion of eGFR measurements that fell within stage 1 was 47\% when computed with CKD-EPI 2009 and 38\% when computed with CKD-EPI 2021 (\tableref{tab:etab5}). When CKD stages were different, CKD-EPI 2021 always classified eGFR measurements into more severe stages, most commonly from CKD stage 1 to CKD stage 2 (55\%) (\tableref{tab:etab6}). 

Across 1,376,397 eGFR measurements corresponding to patients not documented as Black or African American, a total of 167,514 (12\%) measurements had a different CKD stage depending on whether eGFR was computed with CKD-EPI 2009 or CKD-EPI 2021. The proportion of eGFR measurements that fell within stage 1 was 43\% when computed with CKD-EPI 2009 and 50\% when computed with CKD-EPI 2021 (\tableref{tab:etab7}). When CKD stages were different, CKD-EPI 2021 always classified eGFR measurements into less severe CKD stages, most commonly from CKD stage 2 to 1 (58\%) (\tableref{tab:etab8}).

\subsection{Determination of Nephrology Clinics in SHC Data }\label{apd:methods3}

We identified a total of 12 nephrology clinics at SHC, identified by keyword matching according to care site name as recorded in SHC live Epic data. Nephrology clinics were located across the San Francisco Bay area in the following counties: Santa Clara, Alameda, San Mateo, and San Francisco. The majority of nephrology visits (91\%) took place at the SHC Boswell location. 

\subsection{Documentation of Race and Ethnicity in SHC Data }\label{apd:methods4}
The reliability of race and ethnicity data in the electronic health data were unclear. We do not know if these data are self-reported or documented on behalf of patients by providers. Ascertainment of patient self-reported race and ethnicity information was not possible given the size of our patient cohort. 

For 5\% and 3\% of patients in our cohort, there was at least two race or ethnicity categories documented. For each of the eight race and four ethnicity categories recorded at SHC, we recorded a binary indicator and patients could be recorded in multiple categories. Thus, race and ethnicity categories in \tableref{tab:cohortstudy} do not sum to 1. 

\subsection{Time from Nephrology Referral to Visit at Nephrology Clinic }\label{apd:methods5}
We identified 6,828 (64\%) patients that had a recorded nephrology referral prior to their first visit at the SHC nephrology clinic and for whom it was possible to calculate time from referral to visit. For these patients, we identified the post-referral visit (visit closest after referral rate), and calculated time (days) from referral to post-referral visit. In \tableref{tab:etab2}, we computed the proportion of patients with identified referrals during our study period. Years were assigned according to the post-referral visit date. 

We investigated whether there were differences in the time from referral to visit among those who were documented as Black or African American or not Black or African American. Among patients with a visit at an SHC nephrology clinic, 897 were documented as Black or African American, and 12,337 patients were documented as not Black or African American. The proportion of patients with nephrology referrals was 61\% for those documented as Black or African American patients and 57\% for those not documented as Black or African American. In \tableref{tab:etab3}, we calculated yearly statistics for the proportion of patients with a post-referral visit and days from referral to visit at SHC nephrology clinics for those documented as Black or African American or not Black or African American. There were no differences in mean and median days from nephrology referral to visit among these two subgroups. 

\subsection{Possible Seasonality Adjustments }\label{apd:methods6}
While we do not expect CKD to be seasonal, it is possible that rates of nephrology referrals and clinics follow seasonal patterns because they were normalized by overall patient counts at SHC, which include visits due to infectious diseases. To adjust for possible seasonality, we included binary variables for each quarter in an interrupted time series regression sensitivity analysis (\tableref{tab:etab16}). 


\clearpage

\onecolumn
\section{Appendix Tables and Figures}\label{apd:second}

\begin{figure}[h]
 \floatconts
  {fig:efig1}
  {\caption{Quarterly patients seen at SHC during study period (documented as Black or African American)}}
  {\includegraphics[width=0.75\textwidth]{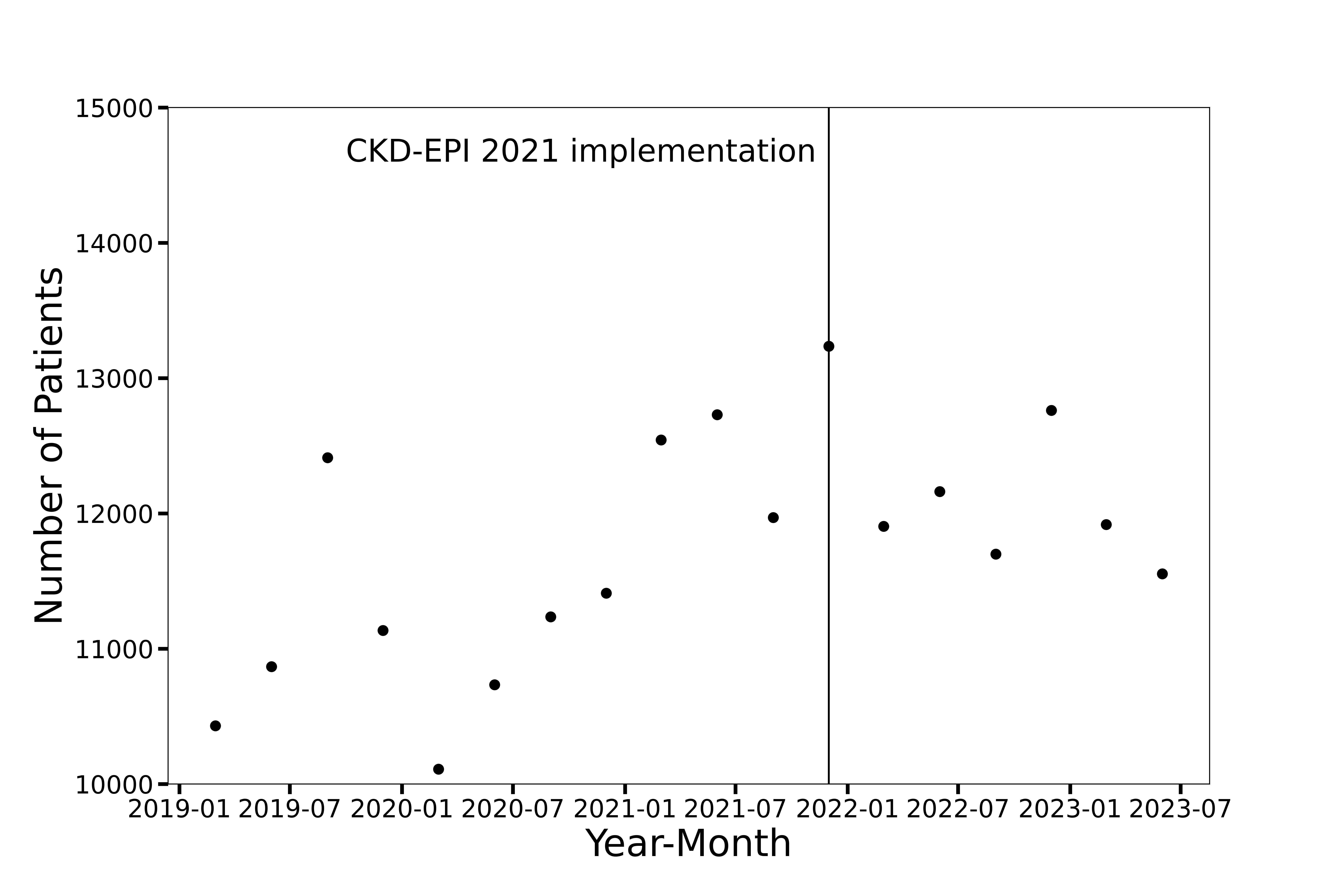}}
\end{figure}

\begin{figure}[h]
\floatconts
  {fig:efig2}
  {\caption{Quarterly patients seen at SHC during study period (not documented as Black or African American)}}
  {\includegraphics[width=0.75\textwidth]{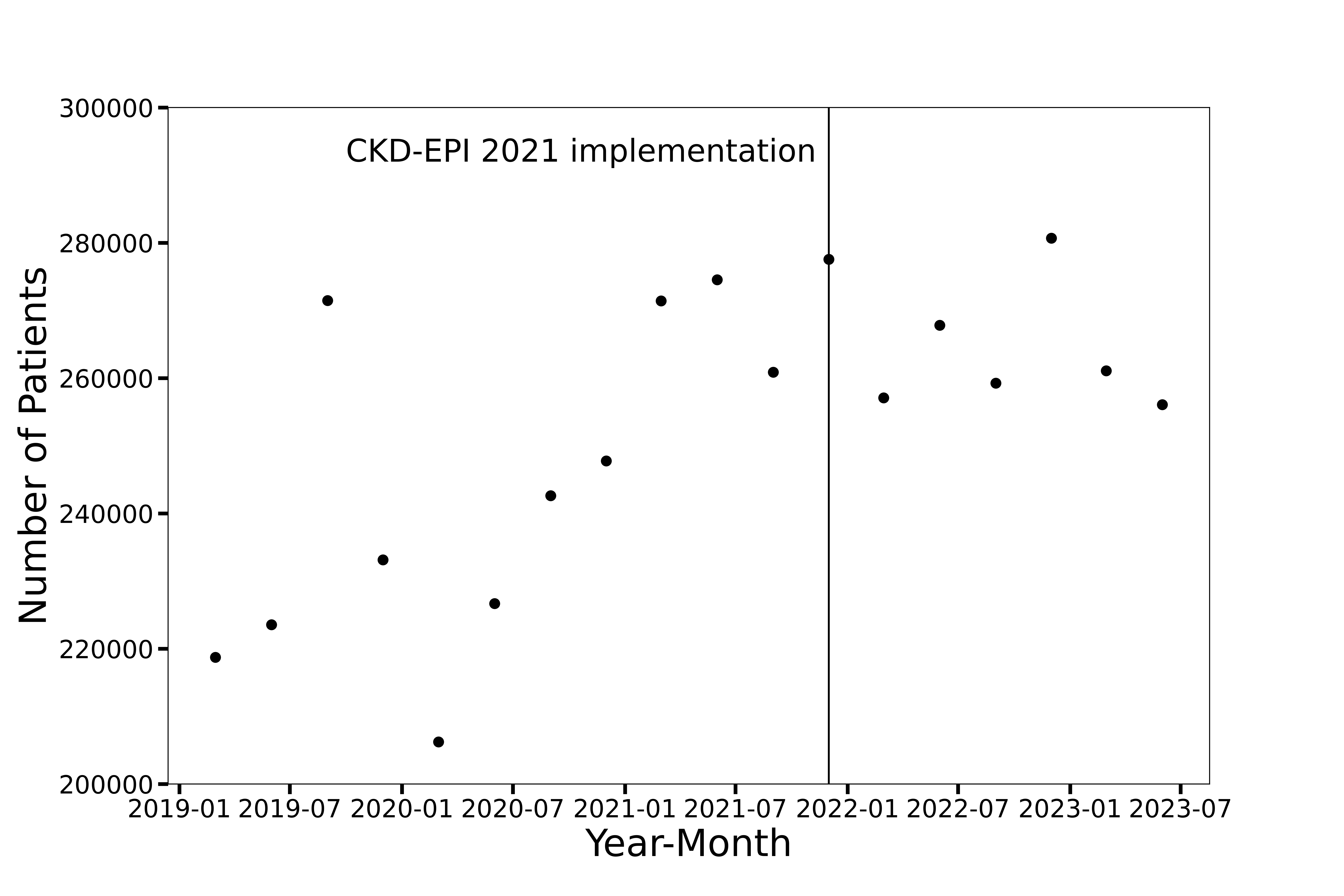}}
\end{figure}

\begin{figure}[h]
\floatconts
  {fig:efig3}
  {\caption{Quarterly patients seen at SHC during study period (overall cohort)}}
  {\includegraphics[width=0.75\textwidth]{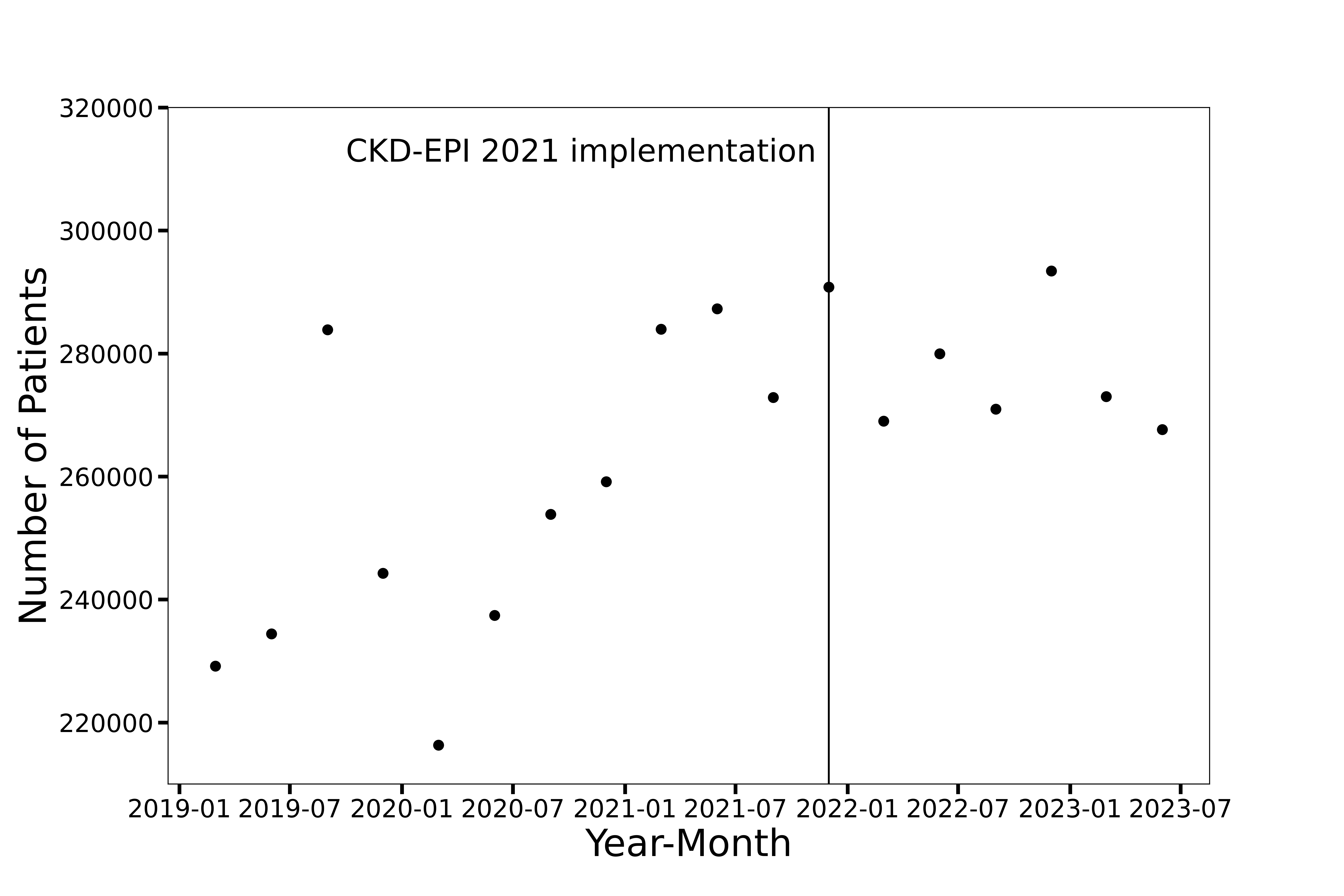}}
\end{figure}

\begin{figure}[h]
\floatconts
  {fig:efig4}
  {\caption{Proportion of visits at SHC using various eGFR equations during study period}}
  {\includegraphics[width=0.75\textwidth]{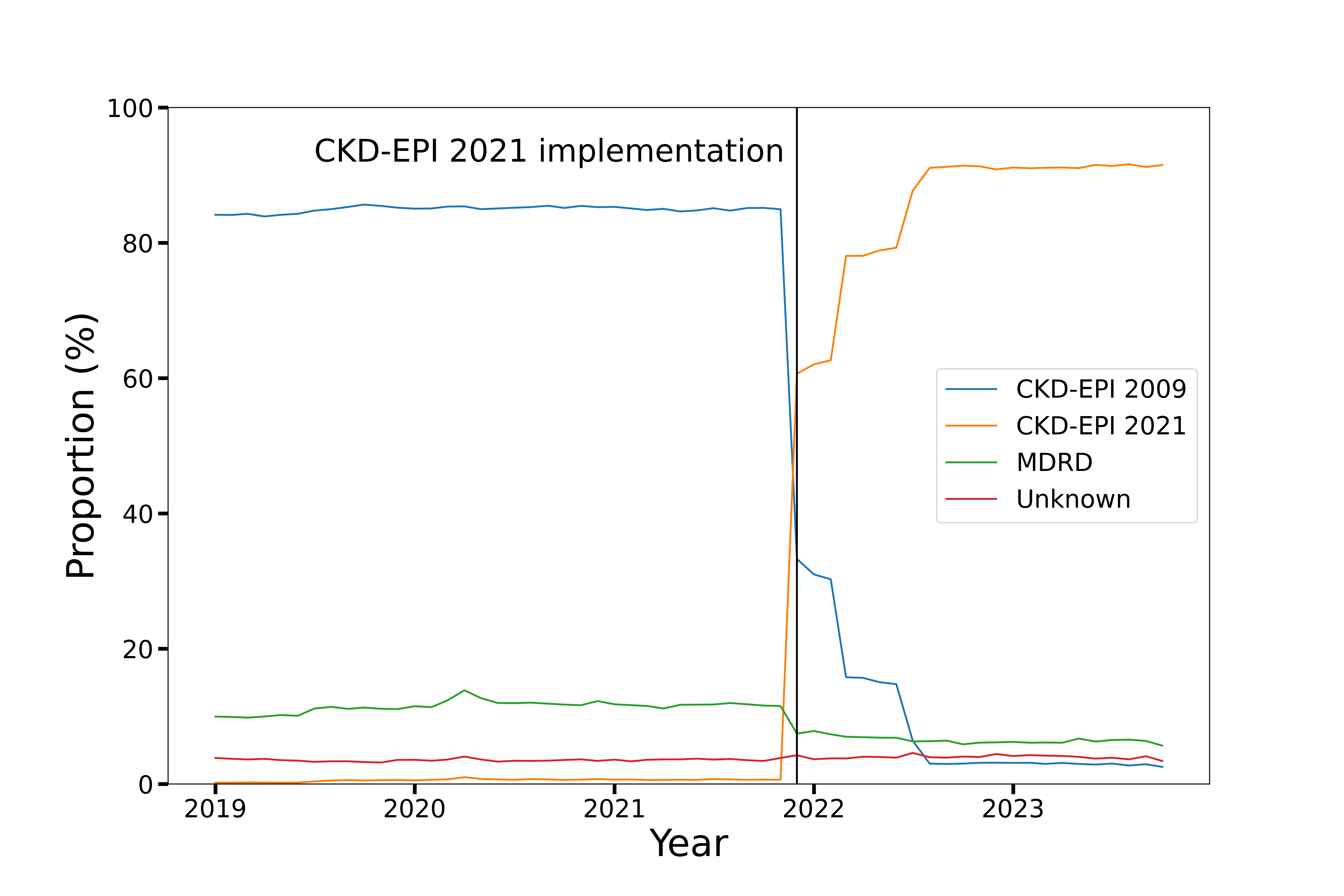}}
\end{figure}

\begin{figure}[h]
\floatconts
  {fig:efig5}
  {\caption{Quarterly median time from nephrology referral to nephrology visit (documented as Black or African American)}}
  {\includegraphics[width=0.75\textwidth]{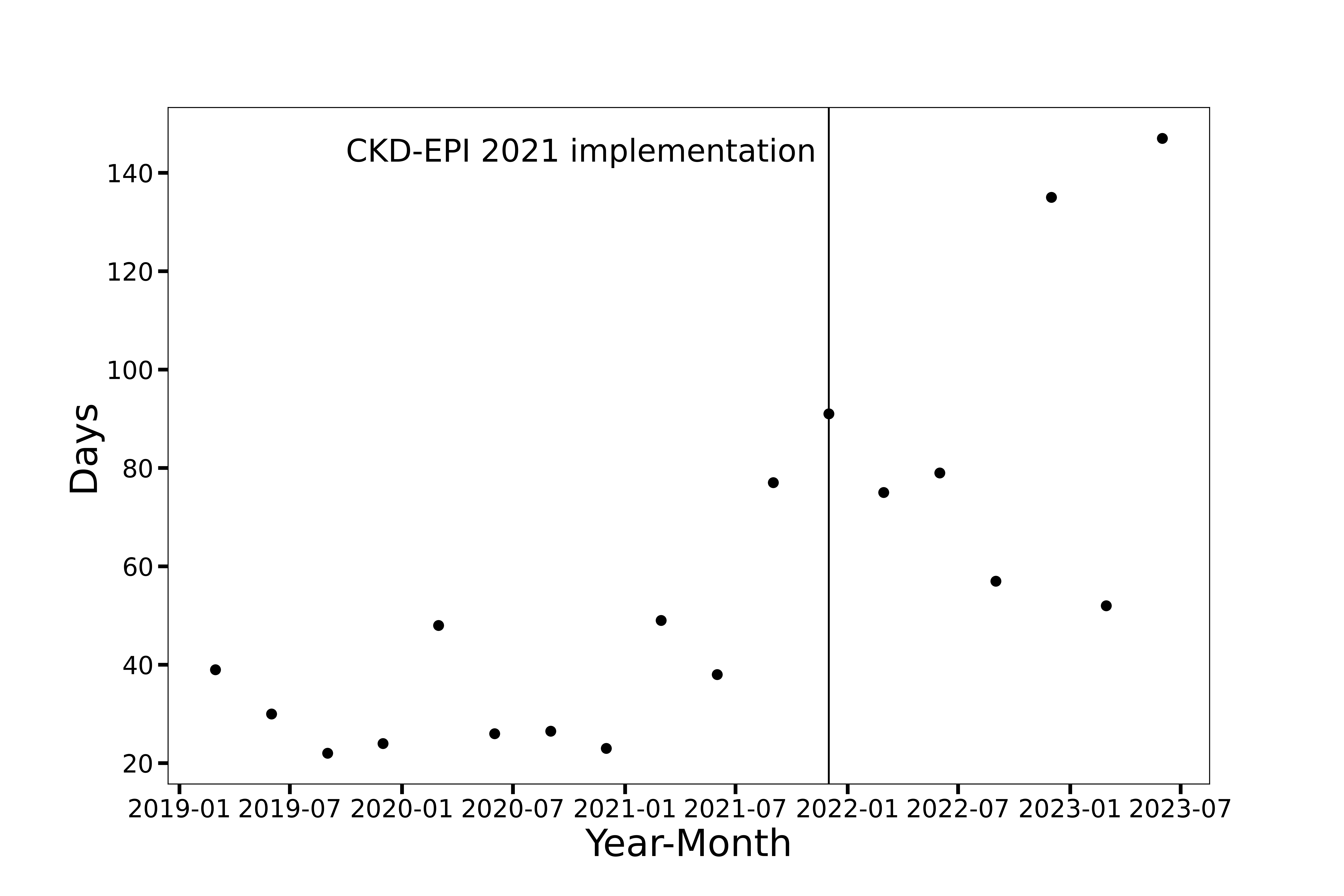}}
\end{figure}

\begin{figure}[h]
\floatconts
  {fig:efig6}
  {\caption{Quarterly median time from nephrology referral to nephrology visit (not documented as Black or African American)}}
  {\includegraphics[width=0.75\textwidth]{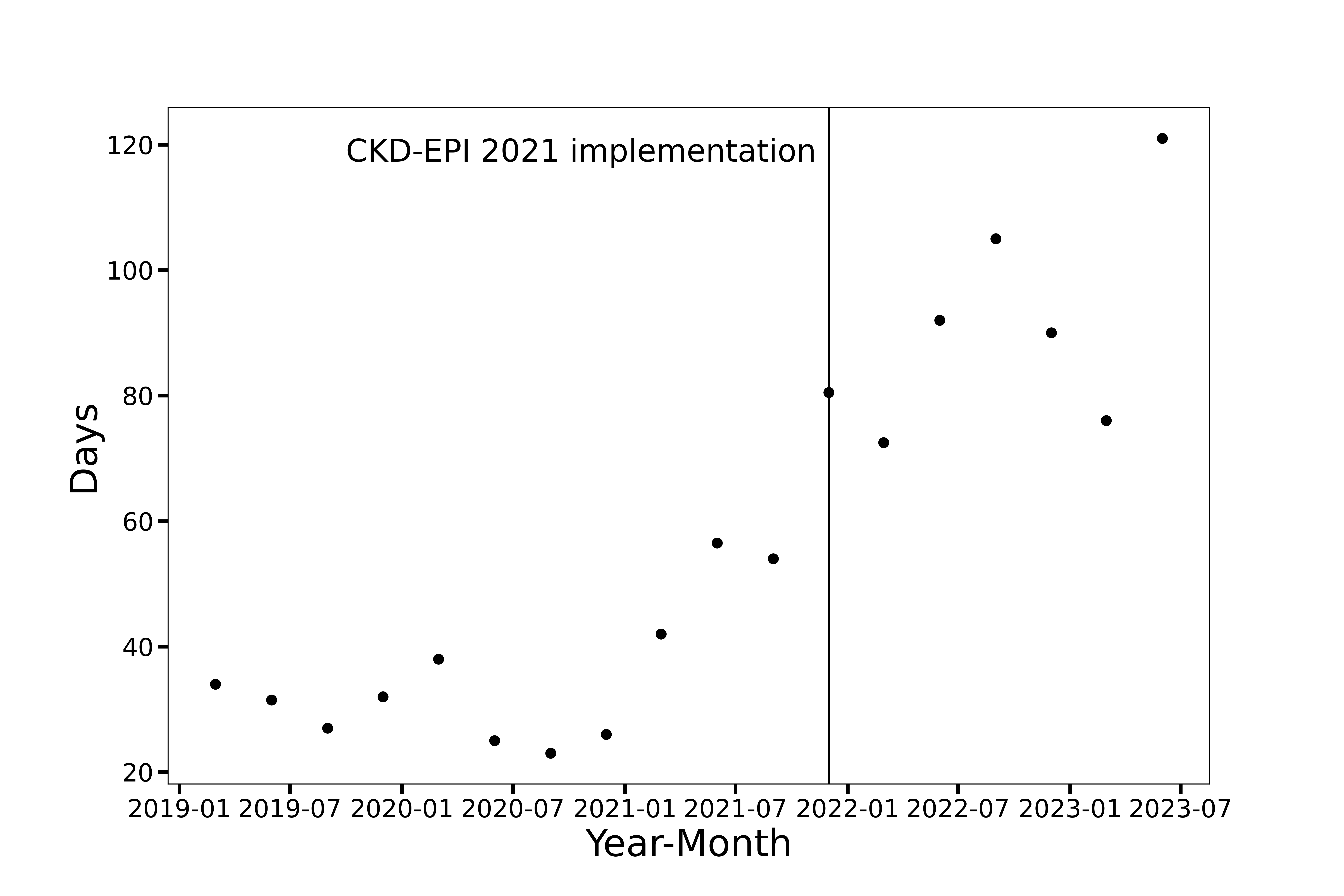}}
\end{figure}

\begin{figure}[h]
\floatconts
  {fig:efig7}
  {\caption{Quarterly median time from nephrology referral to nephrology visit (overall cohort)}}
  {\includegraphics[width=0.75\textwidth]{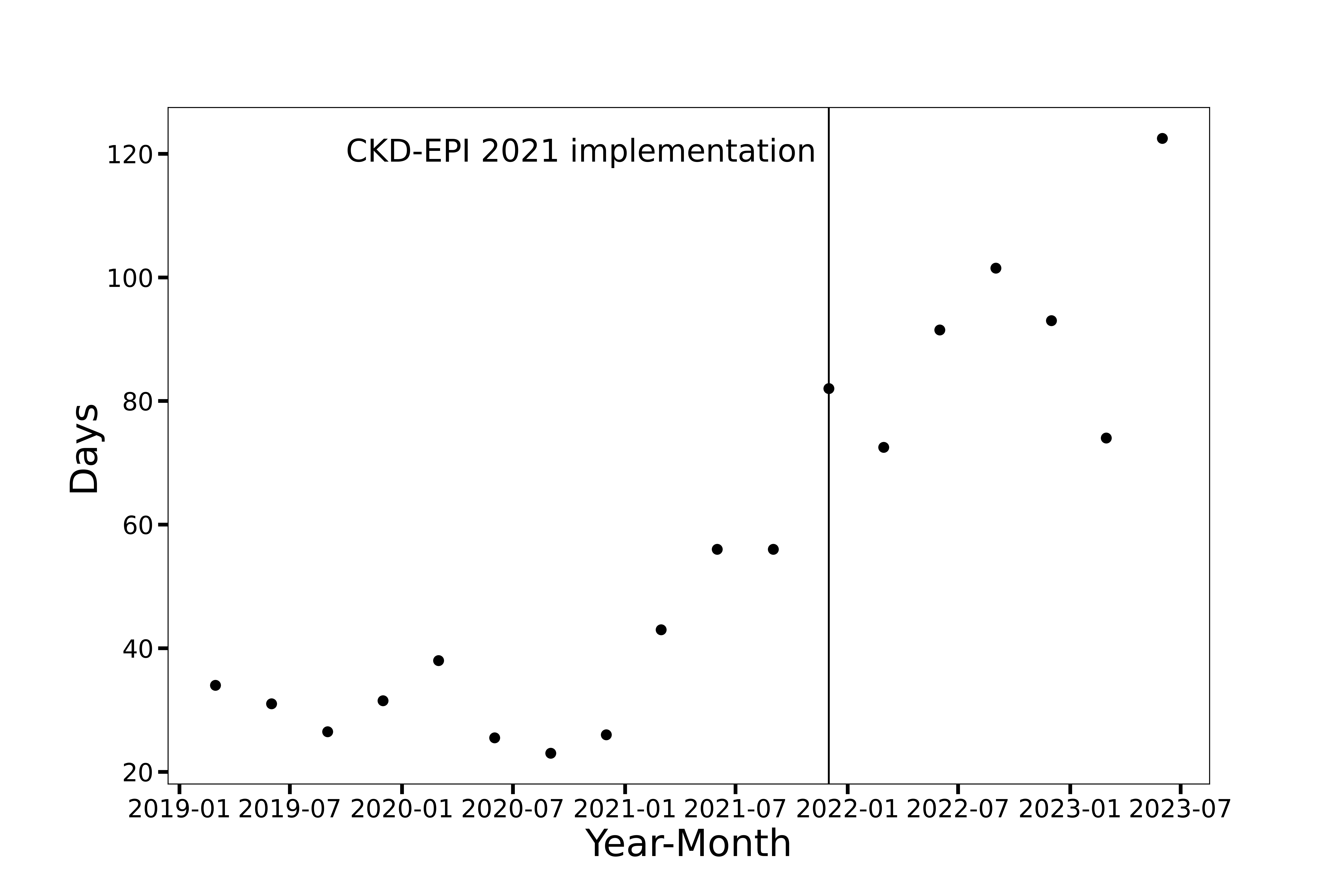}}
\end{figure}

\begin{table*}[h]
    \floatconts {tab:etab1}%
      {\caption{Proportion of visits using various eGFR equations before and after December 1, 2021, SHC eGFR equation usage shift from CKD-EPI 2009 to CKD-EPI 2021}}%
    {%
    \begin{tabular}{ llll}
    \bfseries eGFR Equation  & \bfseries Overall  & \bfseries Before December 1, 2021 &  \bfseries After December 1, 2021 \\
    \hline\abovestrut{2.2ex}
      \rule{0pt}{3ex}CKD-EPI 2009 & 53\% & 86\% & 5\%\\
     \rule{0pt}{3ex}CKD-EPI 2021 & 37\% & 0\% & 87\% \\
    \rule{0pt}{3ex}MDRD & 7\% & 9\% & 4\% \\
      \belowstrut{0.2ex} 
     \rule{0pt}{3ex}Unknown & 3\% & 5\% & 4\% \\ \hline
    \end{tabular}
}
\end{table*}

\begin{table*}[h]
    \floatconts {tab:etab2}%
      {\caption{Proportion of patients with post-referral nephrology visits and wait time (days) across study period}}%
    {%
    \begin{tabular}{ lll}
    \bfseries Year  & \bfseries Proportion with Post-Referral Visit  & \bfseries Average Days (Median) \\
    \hline\abovestrut{2.2ex}
     \rule{0pt}{3ex}2019 & 68\% & 77 (32)\\
    \rule{0pt}{3ex}2020 & 68\% & 74 (27)\\
    \rule{0pt}{3ex}2021 & 68\% & 100 (40)\\
     \rule{0pt}{3ex}2022 & 64\% & 151 (85)\\
     \belowstrut{0.2ex} 
    \rule{0pt}{3ex}2023 & 62\% & 191 (100)\\ \hline
    \end{tabular}
}
\end{table*}

\begin{table*}[h]
    \floatconts {tab:etab3}%
      {\caption{Proportion of patients with post-referral nephrology visits and wait time (days) across study period for those documented as Black or African American and not Black or African American}}%
    {%
    \begin{tabular}{ lllll}
    &  \multicolumn{2}{l}{\bfseries Proportion with Post-Referral Visit} &  \multicolumn{2}{l}{\bfseries Average Days  (Median)} \\
     \cline{2-5}\abovestrut{2.2ex}
     & Black or & Not Black or  & Black or & Not Black or \\
   \bfseries Year & African American & African American & African American & African American \\
        \hline\abovestrut{2.2ex}
     \rule{0pt}{3ex}2019 & 75\% & 67\% & 70 (31) & 77 (33)\\
    \rule{0pt}{3ex}2020 & 73\% & 68\% & 135 (29) & 70 (27)\\
    \rule{0pt}{3ex}2021 & 75\% & 68\% & 169 (49) & 94 (39)\\
     \rule{0pt}{3ex}2022 & 76\% & 63\% & 185 (79) & 148 (85)\\
     \belowstrut{0.2ex} 
    \rule{0pt}{3ex}2023 & 57\% & 63\% & 235 (114) & 188 (97)\\ \hline
    \end{tabular}
}
\end{table*}

\begin{table*}[h]
    \floatconts {tab:etab4}%
      {\caption{Percent changes in eGFR measurements recorded after December 1, 2021, using CKD-EPI 2009 and CKD-EPI 2021 }}%
    {%
    \begin{tabular}{ ll}
    \bfseries Cohort & \bfseries Percent Change\\
    \hline\abovestrut{2.2ex}
     \rule{0pt}{3ex}Black or African American$^*$ & -10\% \\
     \rule{0pt}{3ex}Not Black or African American$^*$ & 5\% \\
     \belowstrut{0.2ex} 
     \rule{0pt}{3ex}Overall cohort & 4\%\\ \hline
    \end{tabular}
}
\begin{tablenotes}\footnotesize
\item[*] $^*$ Documented race
\end{tablenotes}
\end{table*}

\begin{table*}[h]
    \floatconts {tab:etab5}%
      {\caption{CKD stage distributions for eGFR measurements recorded after December 1, 2021, using CKD-EPI 2009 and CKD-EPI 2021 (documented as Black or African American)}}%
    {%
    \begin{tabular}{lll}
    \bfseries CKD Stage & \bfseries CKD-EPI 2009 & \bfseries CKD-EPI 2021 \\
     \hline
     \abovestrut{2.2ex}
     \rule{0pt}{3ex}1 (90+mL/min/1.73m2) & 47\% & 38\% \\
     \rule{0pt}{3ex}2 (60-89mL/min/1.73m2)  & 28\% & 33\% \\
     \rule{0pt}{3ex}3a (45-59mL/min/1.73m2) & 10\% & 12\% \\
     \rule{0pt}{3ex}3b (30-44mL/min/1.73m2) & 7\% & 8\% \\
     \rule{0pt}{3ex}4 (15-29mL/min/1.73m2 & 5\% & 5\% \\ 
     \belowstrut{0.2ex} 
    \rule{0pt}{3ex}5 ($<$15mL/min/1.73m2) & 4\% & 4\% \\ \hline
    \end{tabular}
}
\end{table*}

\begin{table*}[h]
    \floatconts {tab:etab6}%
      {\caption{Proportion of eGFR measurements recorded after December 1, 2021, with different CKD stage assignment when computed by CKD-EPI 2009 and CKD-EPI 2021 (documented as Black or African American)}}%
    {%
    \begin{tabular}{lll}
    \bfseries CKD-EPI 2009 Stage & \bfseries CKD-EPI 2021 Stage & \bfseries Proportion \\
     \hline
     \abovestrut{2.2ex}
     \rule{0pt}{3ex}1 & 2 & 55\% \\
    \rule{0pt}{3ex}2 & 3a & 25\% \\
    \rule{0pt}{3ex}3a & 3b & 12\% \\
     \rule{0pt}{3ex}3b & 4 & 6\% \\
     \belowstrut{0.2ex} 
    \rule{0pt}{3ex}4 & 5 & 2\% \\ \hline
    \end{tabular}
}
\end{table*}

\begin{table*}[h]
    \floatconts {tab:etab7}%
      {\caption{CKD stage distributions for eGFR measurements recorded after December 1, 2021, using CKD-EPI 2009 and CKD-EPI 2021 (not documented as Black or African American)}}%
    {%
    \begin{tabular}{lll}
    \bfseries CKD Stage & \bfseries CKD-EPI 2009 & \bfseries CKD-EPI 2021 \\
    \hline\abovestrut{2.2ex}
     \rule{0pt}{3ex}1 (90+mL/min/1.73m2) & 43\% & 50\% \\
     \rule{0pt}{3ex}2 (60-89mL/min/1.73m2)  & 35\% & 31\% \\
     \rule{0pt}{3ex}3a (45-59mL/min/1.73m2) & 10\% & 9\% \\
     \rule{0pt}{3ex}3b (30-44mL/min/1.73m2) & 7\% & 6\% \\
     \rule{0pt}{3ex}4 (15-29mL/min/1.73m2 & 4\% & 3\% \\ 
     \belowstrut{0.2ex} 
    \rule{0pt}{3ex}5 ($<$15mL/min/1.73m2) & 2\% & 2\% \\ \hline
    \end{tabular}
}
\end{table*}

\begin{table*}[h]
    \floatconts {tab:etab8}%
      {\caption{Proportion of eGFR measurements recorded after December 1, 2021, with different CKD stage assignment when computed by CKD-EPI 2009 and CKD-EPI 2021 (not documented as Black or African American)}}%
    {%
    \begin{tabular}{lll}
    \bfseries CKD-EPI 2009 Stage & \bfseries CKD-EPI 2021 Stage & \bfseries Proportion \\
    \hline\abovestrut{2.2ex}
     \rule{0pt}{3ex}2 & 1 & 58\% \\
     \rule{0pt}{3ex}3a & 2 & 23\% \\
    \rule{0pt}{3ex}3b & 3a & 12\% \\
     \rule{0pt}{3ex}4 & 3b & 5\% \\
     \belowstrut{0.2ex} 
   \rule{0pt}{3ex}5 & 4 & 2\% \\ \hline
    \end{tabular}
}
\end{table*}
\begin{table*}[h]
    \floatconts {tab:etab9}%
      {\caption{CKD stage distributions for eGFR measurements recorded after December 1, 2021, using CKD-EPI 2009 and CKD-EPI 2021 (overall cohort)}}%
    {%
    \begin{tabular}{lll}
    \bfseries CKD Stage & \bfseries CKD-EPI 2009 & \bfseries CKD-EPI 2021 \\
    \hline\abovestrut{2.2ex}
     \rule{0pt}{3ex}1 (90+mL/min/1.73m2) & 43\% & 50\% \\
     \rule{0pt}{3ex}2 (60-89mL/min/1.73m2)  & 35\% & 31\% \\
     \rule{0pt}{3ex}3a (45-59mL/min/1.73m2) & 10\% & 9\% \\
     \rule{0pt}{3ex}3b (30-44mL/min/1.73m2) & 7\% & 6\% \\
     \rule{0pt}{3ex}4 (15-29mL/min/1.73m2 & 4\% & 3\% \\ 
     \belowstrut{0.2ex} 
    \rule{0pt}{3ex}5 ($<$15mL/min/1.73m2) & 2\% & 2\% \\ \hline
    \end{tabular}
}
\end{table*}
\begin{table*}[h]
    \floatconts {tab:etab10}%
      {\caption{Proportion of eGFR measurements recorded after December 1, 2021, with different CKD stage assignment when computed by CKD-EPI 2009 and CKD-EPI 2021 (overall cohort)}}%
    {%
    \begin{tabular}{lll}
    \bfseries CKD-EPI 2009 Stage & \bfseries CKD-EPI 2021 Stage & \bfseries Proportion \\
    \hline\abovestrut{2.2ex}
     \rule{0pt}{3ex}2 & 1 & 54\% \\
     \rule{0pt}{3ex}3a & 2 & 22\% \\
    \rule{0pt}{3ex}3b & 3a & 12\% \\
     \rule{0pt}{3ex}4 & 3b & 5\% \\
     \rule{0pt}{3ex}1 & 2 & 4\% \\
     \rule{0pt}{3ex}2 & 3a & 2\% \\
     \rule{0pt}{3ex}5 & 4 & 1\% \\
     \rule{0pt}{3ex}3a & 3b & $<$1\% \\
    \rule{0pt}{3ex}3b & 4 & $<$1\% \\
     \belowstrut{0.2ex} 
    \rule{0pt}{3ex}4 & 5 & $<$1\% \\ \hline
    \end{tabular}
}
\end{table*}

\begin{table*}[h]
    \floatconts {tab:etab11}%
      {\caption{Rate ratios (RR) of implementing CKD-EPI 2021 and estimated quarterly rates of nephrology referrals and visits under CKD-EPI 2021 and race-adjusted eGFR with confidence intervals (95\% CIs) from interrupted time series regression for overall cohort and subgroups (Black or African American, not Black or African American) under 3 month temporary lag period}}%
    {%
    \begin{tabular}{ llll}
    \bfseries  & \bfseries   & \bfseries Black or &  \bfseries Not Black or  \\
    \bfseries  & \bfseries Overall & \bfseries African American$^*$ &  \bfseries African American$^*$ \\
    \hline\abovestrut{2.2ex}
    \textbf{Nephrology referral rates} & & & \\ 
    
    \rule{0pt}{3ex}Adjusted RR & 0.94 (0.85, 1.03) & 0.80 (0.56, 1.15) & 0.95 (0.86, 1.05)\\
    (CKD-EPI 2021) & & & \\
    \rule{0pt}{3ex}Estimated rate per 10,000 people & 21 (20, 22) & 33 (29, 39) & 20 (19, 21)\\
     (CKD-EPI 2021) & & & \\
    \rule{0pt}{3ex}Estimated rate per 10,000 people & 22 (20, 24) & 41 (30, 57) & 21 (19, 23)\\
     (race adjusted) & & & \\
    \rule{0pt}{4ex}\textbf{Nephrology visit rates} & & & \\ 
    \rule{0pt}{3ex}Adjusted RR & 1.04 (1.00, 1.09) & 1.04 (0.89, 1.22) & 1.04 (1.00, 1.09)\\
    (CKD-EPI 2021) & & & \\
    \rule{0pt}{3ex}Estimated rate per 10,000 people & 111 (109, 112) & 189 (177, 201) & 102 (100, 104)\\
    (CKD-EPI 2021) & & & \\
    \rule{0pt}{3ex}Estimated rate per 10,000 people & 106 (102, 111) & 181 (157, 210) & 99 (95, 103) \\
    \belowstrut{0.2ex}
    (race adjusted) & & & \\\hline
    \end{tabular}
}
\begin{tablenotes}\footnotesize
\item[*] $^*$ Documented race
\end{tablenotes}
\end{table*}

\begin{table*}[h]
    \floatconts {tab:etab12}%
      {\caption{Rate ratios (RR) of implementing CKD-EPI 2021 and estimated quarterly rates of nephrology referrals and visits under CKD-EPI 2021 and race-adjusted eGFR with confidence intervals (95\% CIs) from interrupted time series regression for overall cohort and subgroups (Black or African American, not Black or African American) under 6 month temporary lag period}}%
    {%
    \begin{tabular}{ llll}
    \bfseries  & \bfseries  & \bfseries Black or &  \bfseries Not Black or  \\
    \bfseries  & \bfseries Overall & \bfseries African American$^*$ &  \bfseries African American$^*$ \\
    \hline\abovestrut{2.2ex}
    \textbf{Nephrology referral rates} & & & \\
    \rule{0pt}{3ex}Adjusted RR & 0.91 (0.82, 1.01) & 0.81 (0.56, 1.19) & 0.92 (0.83, 1.02)\\
    (CKD-EPI 2021) & & & \\
    \rule{0pt}{3ex}Estimated rate per 10,000 people & 21 (20, 22) & 34 (29, 39) & 20 (19, 21)\\
    (CKD-EPI 2021) & & & \\
    \rule{0pt}{3ex}Estimated rate per 10,000 people & 23 (21, 25) & 40 (29, 56) & 22 (20, 24)\\
    (race adjusted) & & & \\
    \rule{0pt}{4ex}\textbf{Nephrology visit rates} & & & \\
    \rule{0pt}{3ex}Adjusted RR & 1.01 (0.97, 1.06) & 0.99 (0.84, 1.17) & 1.01 (0.97, 1.06)\\
    (CKD-EPI 2021) & & & \\
    \rule{0pt}{3ex}Estimated rate per 10,000 people & 110 (108, 112) & 188 (176, 201) & 102 (100, 104)\\
    (CKD-EPI 2021) & & & \\
    \rule{0pt}{3ex}Estimated rate per 10,000 people & 109 (105, 114) & 189 (164, 219) & 101 (97, 105) \\ 
    \belowstrut{0.2ex}
    (race adjusted) & & & \\\hline
    \end{tabular}
}
\begin{tablenotes}\footnotesize
\item[*] $^*$ Documented race
\end{tablenotes}
\end{table*}

\begin{table*}[h]
    \floatconts {tab:etab13}%
      {\caption{Rate ratios (RR) of implementing CKD-EPI 2021 and estimated quarterly rates of nephrology referrals and visits under CKD-EPI 2021 and race-adjusted eGFR with confidence intervals (95\% CIs) from interrupted time series regression for overall cohort and subgroups (Black or African American, not Black or African American) under 12 month temporary lag period}}%
    {%
    \begin{tabular}{llll}

    \bfseries  & \bfseries  & \bfseries Black or &  \bfseries Not Black or  \\
    \bfseries  & \bfseries Overall & \bfseries African American$^*$ &  \bfseries African American$^*$ \\
    \hline\abovestrut{2.2ex}
    \textbf{Nephrology referral rates} & & & \\ 
    
    \rule{0pt}{3ex}Adjusted RR & 0.96 (0.87, 1.07) & 0.91 (0.61, 1.34) & 0.97 (0.87, 1.08)\\
    (CKD-EPI 2021) & & & \\
    \rule{0pt}{3ex}Estimated rate per 10,000 people & 21 (20, 22) & 34 (29, 40) & 20 (19, 21)\\
    (CKD-EPI 2021) & & & \\
    \rule{0pt}{3ex}Estimated rate per 10,000 people & 21 (20, 23) & 36 (27, 49) & 21 (19, 23)\\
    (race adjusted) & & & \\
    \rule{0pt}{4ex}\textbf{Nephrology visit rates} & & & \\ 
    \rule{0pt}{3ex}Adjusted RR & 1.03 (0.97, 1.07) & 1.00 (0.85, 1.20) & 1.02 (0.97, 1.07)\\
    (CKD-EPI 2021) & & & \\
   \rule{0pt}{3ex}Estimated rate per 10,000 people & 110 (108, 112) & 188 (176, 201) & 102 (100, 104)\\
    (CKD-EPI 2021) & & & \\
    \rule{0pt}{3ex}Estimated rate per 10,000 people  & 108 (104, 113) & 187 (165, 214) & 100 (97, 104)\\
    \belowstrut{0.2ex}
    (race-adjusted) & & & \\\hline
    \end{tabular}
}
\begin{tablenotes}\footnotesize
\item[*] $^*$ Documented race
\end{tablenotes}
\end{table*}

\begin{table*}[h]
    \floatconts {tab:etab14}%
      {\caption{Rate ratios (RR) of implementing CKD-EPI 2021 and estimated quarterly rates of nephrology referrals and visits under CKD-EPI 2021 and race-adjusted eGFR with confidence intervals (95\% CIs) from interrupted time series regression for overall cohort and subgroups (Black or African American, not Black or African American) under alternative impact model: immediate slope and level change}}%
    {%
    \begin{tabular}{ llll}
    \bfseries  & \bfseries   & \bfseries Black or &  \bfseries Not Black or  \\
    \bfseries  & \bfseries Overall & \bfseries African American$^*$ &  \bfseries African American$^*$ \\
    \hline\abovestrut{2.2ex}
    \textbf{Nephrology referral rates} & & & \\ 
    
    \rule{0pt}{3ex}Adjusted RR & 0.90 (0.82, 0.99) & 0.81 (0.57, 1.17) & 0.91 (0.82, 1.00)\\
    (CKD-EPI 2021 immediate) & & & \\
    \rule{0pt}{3ex}Adjusted RR & 1.02 (1.00, 1.04) & 1.03 (0.96, 1.10) & 1.02 (0.99, 1.03)\\
    (CKD-EPI 2021 gradual) & & & \\
    \rule{0pt}{3ex}Estimated rate per 10,000 people & 21 (20, 22) & 34 (28, 40) & 20 (19, 21)\\
    (CKD-EPI 2021) & & & \\
    \rule{0pt}{3ex}Estimated rate per 10,000 people & 22 (19, 24) & 37 (25, 53) & 21 (19, 23)\\
    (race adjusted) & & & \\
    \rule{0pt}{4ex}\textbf{Nephrology visit rates} & & & \\
    \rule{0pt}{3ex}Adjusted RR & 1.03 (0.98, 1.07) & 1.02 (0.87, 1.19) & 1.03 (0.98, 1.07)\\
    (CKD-EPI 2021 immediate) & & & \\
    \rule{0pt}{3ex}Adjusted RR  & 1.01 (0.99, 1.01) & 1.01 (0.98, 1.04) & 1.01 (0.99, 1.01)\\
    (CKD-EPI 2021 gradual) & & & \\
    \rule{0pt}{3ex}Estimated rate per 10,000 people & 110 (108, 113) & 188 (174, 204) & 102 (100, 105)\\
    (CKD-EPI 2021) & & & \\
    \rule{0pt}{3ex}Estimated rate per 10,000 people & 105 (100, 110) & 179 (152, 212) & 98 (93, 102) \\
    \belowstrut{0.2ex}
    (race adjusted)\\\hline
    \end{tabular}
}
\begin{tablenotes}\footnotesize
\item[*] $^*$ Documented race
\end{tablenotes}
\end{table*}

\begin{table*}[h]
    \floatconts {tab:etab15}%
      {\caption{Rate ratios (RR) of implementing CKD-EPI 2021 and estimated quarterly rates of nephrology referrals and visits under CKD-EPI 2021 and race-adjusted eGFR with confidence intervals (95\% CIs) from interrupted time series regression for overall cohort and subgroups (Black or African American, not Black or African American) under alternative time-varying covariate (active providers at SHC nephrology clinics)}}%
    {%
    \begin{tabular}{ llll}
    \bfseries  & \bfseries  & \bfseries Black or &  \bfseries Not Black or  \\
    \bfseries  & \bfseries Overall & \bfseries African American$^*$ &  \bfseries African American$^*$ \\
    \hline\abovestrut{2.2ex}
    \textbf{Nephrology referral rates} & & & \\ 
    \rule{0pt}{3ex}Adjusted RR & 0.91 (0.83, 1.01) & 0.79 (0.55, 1.12) & 0.92 (0.84, 1.02)\\
    (CKD-EPI 2021) & & & \\
    \rule{0pt}{3ex}Estimated rate per 10,000 people & 21 (20, 22) & 34 (29, 39) & 20 (19, 21)\\
    (CKD-EPI 2021) & & & \\
    \rule{0pt}{3ex}Estimated rate per 10,000 people & 22 (21, 24) & 41 (31, 54) & 22 (20, 23)\\
    (race adjusted) & & & \\
    \rule{0pt}{4ex}\textbf{Nephrology visit rates} & & & \\ 
    \rule{0pt}{3ex}Adjusted RR & 1.01 (0.96, 1.05) & 0.96 (0.82, 1.12) & 1.03 (0.98, 1.08)\\
    (CKD-EPI 2021) & & & \\
    \rule{0pt}{3ex}Estimated rate per 10,000 people & 110 (109, 112) & 188 (177, 200) & 102 (100, 104)\\
    (CKD-EPI 2021) & & & \\
    \rule{0pt}{3ex}Estimated rate per 10,000 people  & 110 (106, 114) & 194 (171, 220) & 101 (98, 105) \\
    \belowstrut{0.2ex}
    (race adjusted) & & & \\\hline
    \end{tabular}
}
\begin{tablenotes}\footnotesize
\item[*] $^*$ Documented race
\end{tablenotes}
\end{table*}

\begin{table*}[h]
    \floatconts {tab:etab16}%
      {\caption{Rate ratios (RR) of implementing CKD-EPI 2021 and estimated quarterly rates of nephrology referrals and visits under CKD-EPI 2021 and race-adjusted eGFR with confidence intervals (95\% CIs) from interrupted time series regression for overall cohort and subgroups (Black or African American, not Black or African American) after adjusting for seasonality}}%
    {%
    \begin{tabular}{ llll}
    \bfseries  & \bfseries   & \bfseries Black or &  \bfseries Not Black or  \\
    \bfseries  & \bfseries Overall & \bfseries African American$^*$ &  \bfseries African American$^*$ \\
    \hline\abovestrut{2.2ex}
    \textbf{Nephrology referral rates} & & & \\ 
    \rule{0pt}{3ex}Adjusted RR & 0.93 (0.83, 1.04) & 0.84 (0.56, 1.25) & 0.94 (0.84, 1.05)\\
    (CKD-EPI 2021)  & & & \\
    \rule{0pt}{3ex}Estimated rate per 10,000 people & 21 (20, 22) & 34 (28, 41) & 20 (19, 21)\\
     (CKD-EPI 2021)  & & & \\
    \rule{0pt}{3ex}Estimated rate per 10,000 people & 22 (20, 24) & 38 (27, 55) &21 (19, 23)\\
    (race adjusted)  & & & \\
    \rule{0pt}{4ex}\textbf{Nephrology visit rates} & & & \\ 
    \rule{0pt}{3ex}Adjusted RR  & 1.00 (0.95, 1.05) & 0.95 (0.80, 1.14) & 1.01 (0.96, 1.06)\\
    (CKD-EPI 2021)  & & & \\
    \rule{0pt}{3ex}Estimated rate per 10,000 people & 110 (108, 113) & 189 (173, 205) & 102 (100, 105)\\
    (CKD-EPI 2021)  & & & \\
    \rule{0pt}{3ex}Estimated rate per 10,000 people & 110 (106, 115) & 196 (168, 228) & 102 (97, 106) \\
     \belowstrut{0.2ex}
    (race-adjusted)  & & & \\ \hline
    \end{tabular}
}
\begin{tablenotes}\footnotesize
\item[*] $^*$ Documented race
\end{tablenotes}
\end{table*}

\begin{figure}[h]
\floatconts
  {fig:efig8}
  {\caption{Quarterly average age of patients seen at SHC (overall cohort)}}
  {\includegraphics[width=0.75\textwidth]{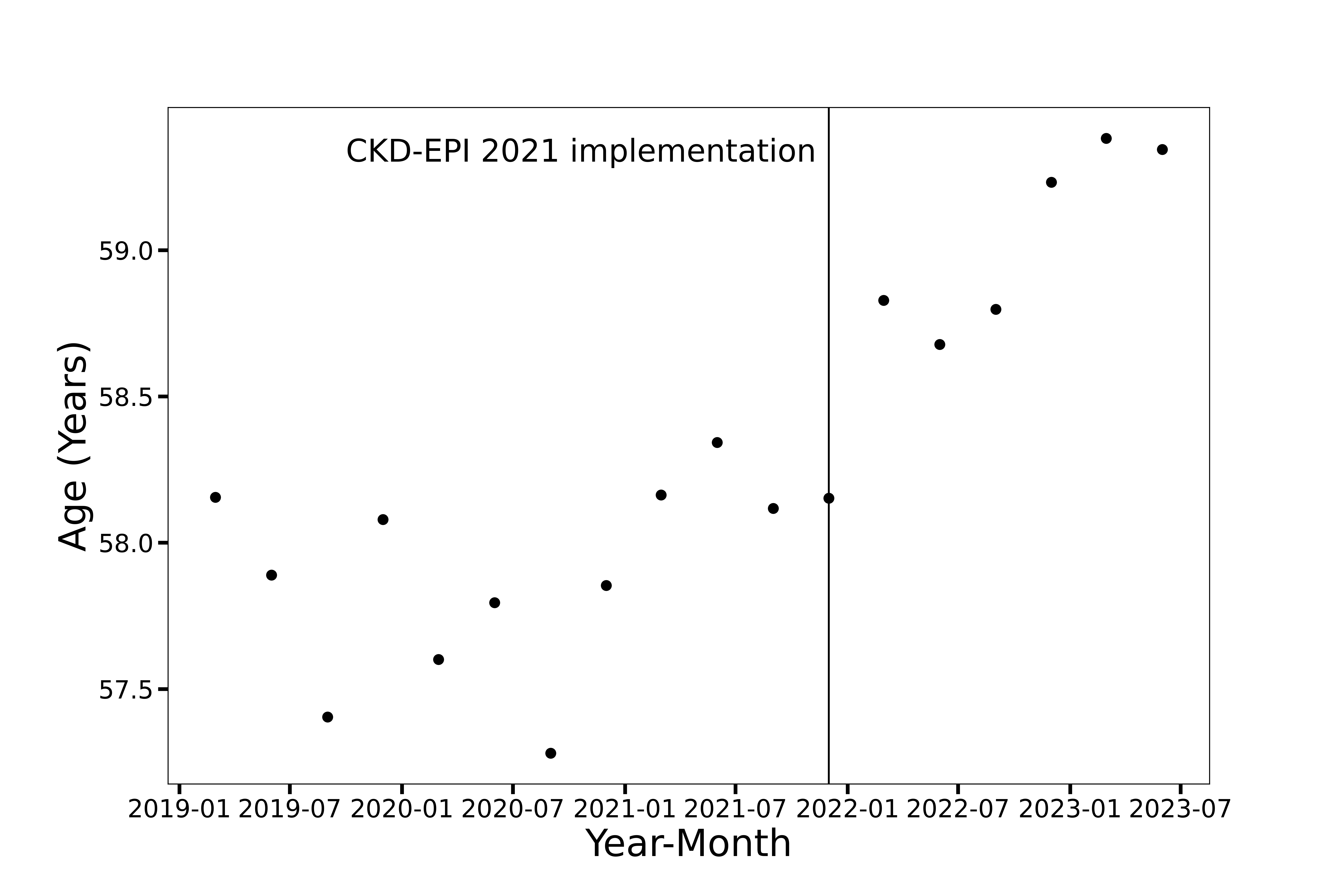}}
\end{figure}

\begin{figure}[h]
\floatconts
  {fig:efig9}
  {\caption{Quarterly proportion of documented female patients at SHC (overall cohort)}}
  {\includegraphics[width=0.75\textwidth]{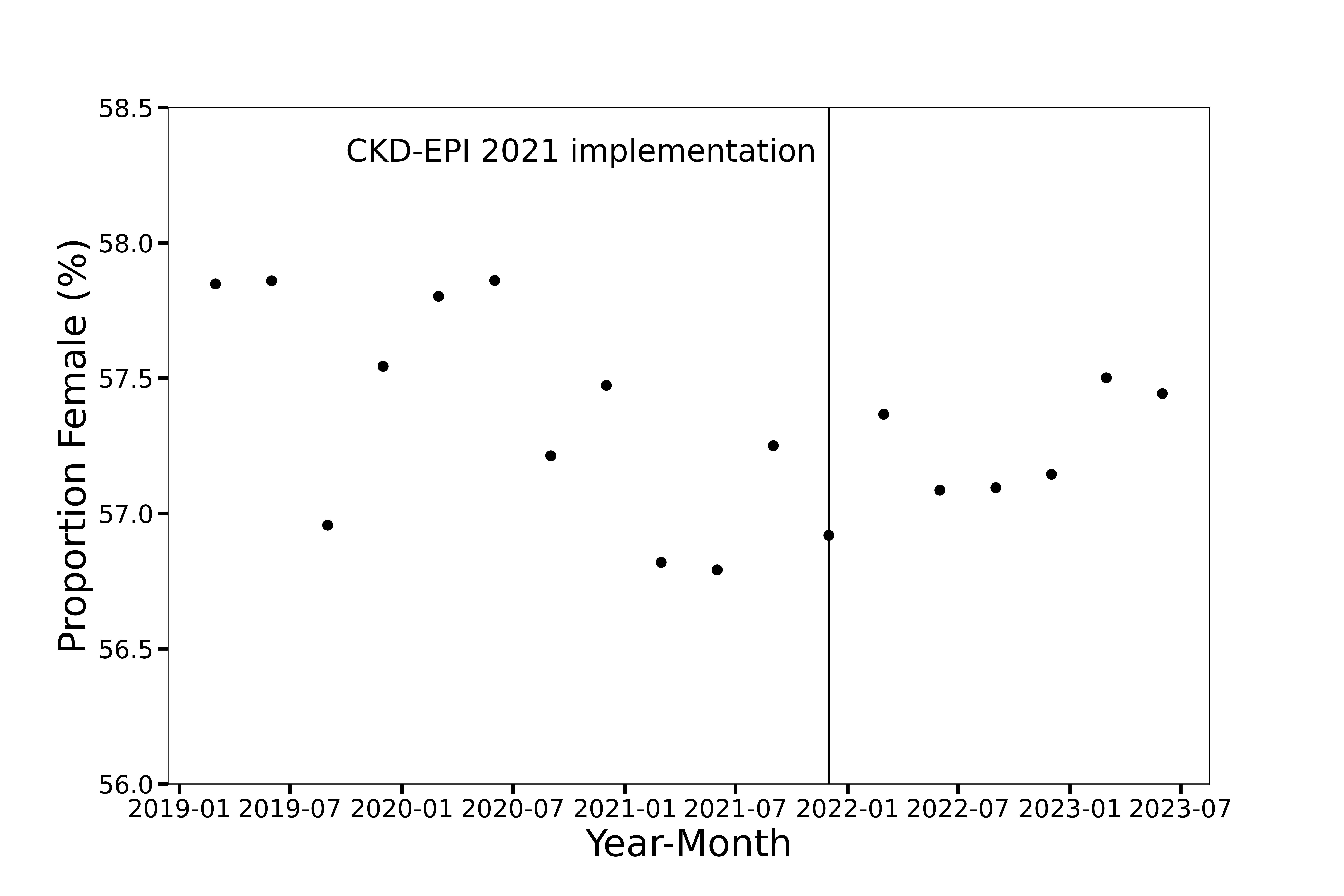}}
\end{figure}

\begin{figure}[h]
\floatconts
  {fig:efig10}
  {\caption{Quarterly count of patients with recorded diabetes diagnosis at SHC (overall cohort)}}
  {\includegraphics[width=0.75\textwidth]{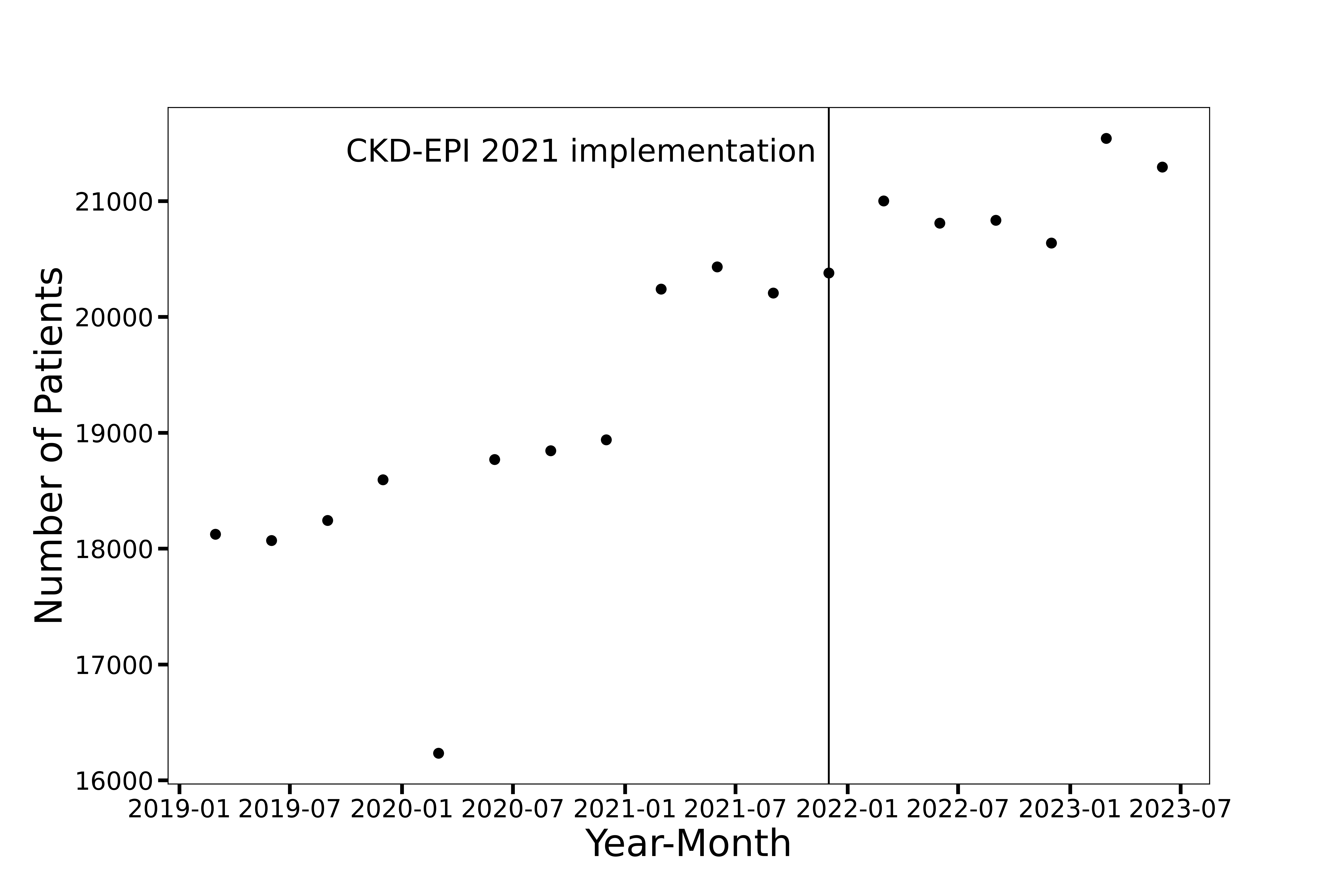}}
\end{figure}

\begin{figure}[h]
\floatconts
  {fig:efig11}
  {\caption{Quarterly count of patients with recorded hypertension diagnosis at SHC (overall cohort)}}
  {\includegraphics[width=0.75\textwidth]{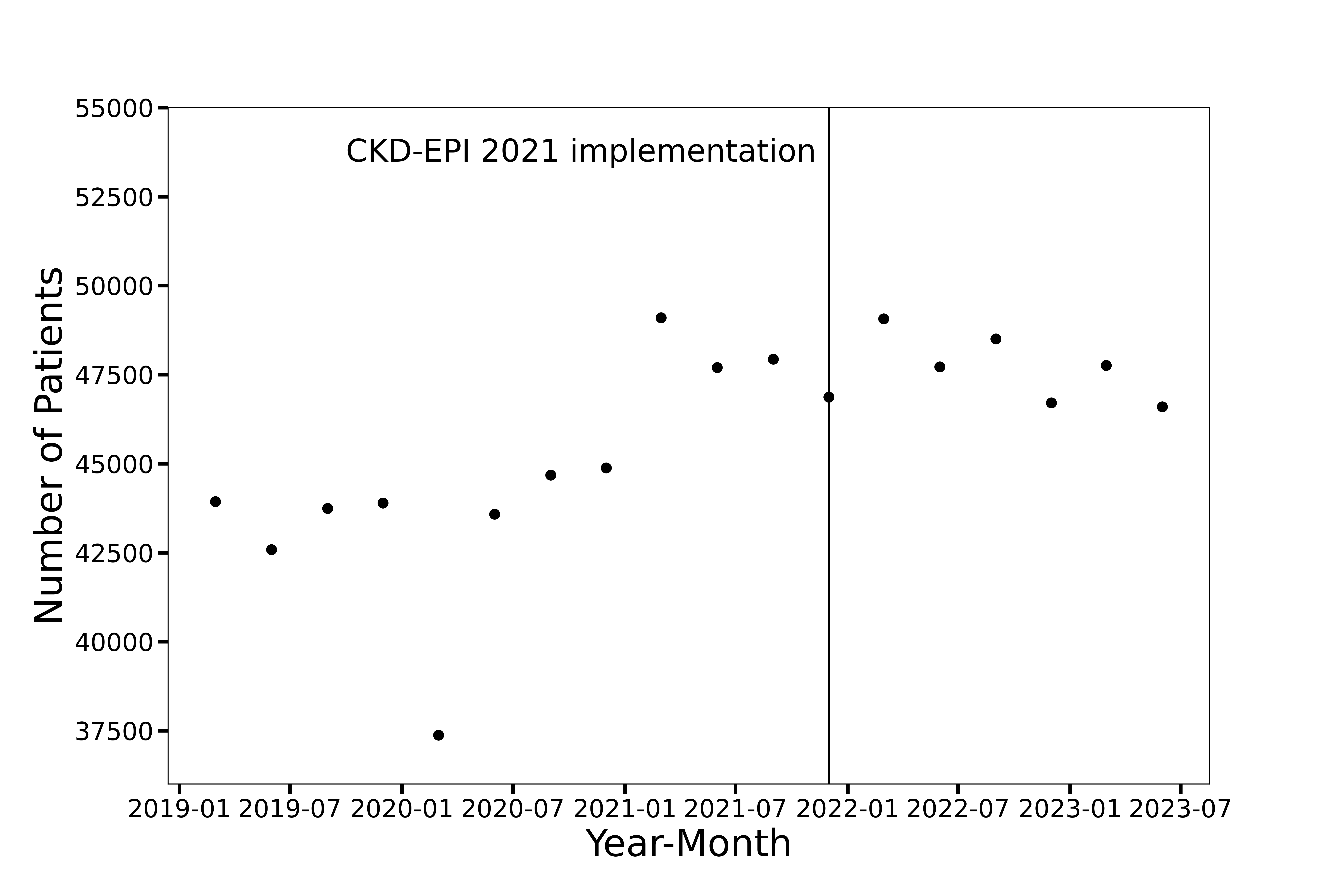}}
\end{figure}

\end{document}